\documentclass[twocolumn]{autart}    

\usepackage[T1]{fontenc}
\usepackage{amssymb,latexsym,amsfonts,amsmath}

\usepackage{wrapfig} 
\newenvironment{authorbio}[2][]{%
	\par\addvspace{1em}%
	\noindent
	\if\relax\detokenize{#1}\relax 
	\else
	\begin{wrapfigure}{l}{0.8in} 
		\vspace{-5pt} 
		\includegraphics[width=1in,height=1.25in,clip,keepaspectratio]{#1}
	\end{wrapfigure}%
	\fi
	\noindent\textbf{#2} 
}{\par\addvspace{2em}}

\usepackage{mdframed,lipsum}

\counterwithin{algorithm}{section}

\usepackage{graphicx}
\usepackage{tikz}
\usetikzlibrary{calc,shapes,arrows,positioning}
\usepackage{pgfplots}

\usepackage{algorithm}
\usepackage{algorithmic}

\usepackage{mathrsfs}

\usepackage{paralist}
\usepackage{dsfont}
\usepackage{eso-pic}
\usepackage{epsfig}
\usepackage{mathtools}
\usepackage{epstopdf}
\usepackage{xfrac}
\usepackage[nocomma]{optidef}

\usepackage{subfig}
\usepackage{multirow}

\usepackage{enumitem}

\usepackage{natbib}

\usepackage{xspace}

\usepackage{xcolor}
\definecolor{lightblue}{rgb}{0.30, 0.75, 0.93}
\definecolor{lightblue}{rgb}{0.30, 0.75, 0.93}
\definecolor{mycolor}{rgb}{0, 0.45, 0.75}
\definecolor{mycolor1}{rgb}{0.39, 0.83, 0.07}
\definecolor{fluorescentpink}{rgb}{1.0, 0.08, 0.58}
\definecolor{royalblue(web)}{rgb}{0.25, 0.41, 0.88}
\definecolor{vividcerise}{rgb}{0.85, 0.11, 0.51}
\definecolor{tangelo}{rgb}{0.98, 0.3, 0.0}
\definecolor{persiangreen}{rgb}{0.0, 0.65, 0.58}
\definecolor{lemon}{rgb}{1.0, 0.97, 0.0}

\DeclareRobustCommand\sampleline[1]{%
	\tikz\draw[#1] (0,0) -- (2em,0);%
}

\DeclareRobustCommand{\legendsquare}[1]{%
	\textcolor{#1}{\rule{2.5ex}{1.5ex}}%
}

\DeclareRobustCommand{\vectorarrow}[1]{%
	\tikz\draw[#1, -{To[width=1.2mm,length=1.5mm]}] (0,0) -- (1.8em,0);%
}%

\DeclareRobustCommand{\legendrect}[1]{%
	\tikz\draw[#1] (0,0) rectangle (2.5ex,1.7ex);%
}%
\usepackage{caption} 
\usepackage{booktabs}
\usepackage {url}
\usepackage{IEEEtrantools}
\usetikzlibrary{arrows,calc,decorations.markings,math,arrows.meta}
\newtheorem{theorem}{Theorem}[section]
\newtheorem{lemma}[theorem]{Lemma}
\newtheorem{proposition}[theorem]{Proposition}

\newtheorem{definition}[theorem]{Definition}

\newtheorem{remark}[theorem]{Remark}
\newtheorem{assumption}[theorem]{Assumption}
\numberwithin{equation}{section}

\usepackage[many]{tcolorbox}
\usetikzlibrary{calc}
\tcbuselibrary{skins}

\newtcolorbox{resp}[1][]{%
	enhanced jigsaw,%
	colback=gray!5!white,%
	colframe=gray!80!black,%
	size=small,%
	boxrule=1pt,%
	halign title=flush center,%
	coltitle=black,%
	breakable,%
	drop shadow=black!50!white,%
	attach boxed title to top left={xshift=1cm,yshift=-\tcboxedtitleheight/2,yshifttext=-\tcboxedtitleheight/2},%
	minipage boxed title=3cm,%
	boxed title style={%
		colback=white,%
		size=fbox,%
		boxrule=1pt,%
		boxsep=2pt,%
		underlay={%
			\coordinate (dotA) at ($(interior.west) + (-0.5pt,0)$);
			\coordinate (dotB) at ($(interior.east) + (0.5pt,0)$);
			\begin{scope}[gray!80!black]
				\fill (dotA) circle (2pt);
				\fill (dotB) circle (2pt);
			\end{scope}
		}%
	},%
	#1%
}

\newcommand{\R}{{\mathbb{R}}}

\newcommand{\N}{{\mathbb{N}}}

\newcommand{\ie}{{\it i.e.}}
\newcommand{\eg}{{\it e.g.}}

\newcommand{\D}{\displaystyle}

\definecolor{gdash}{rgb}{0.10,0.82,0.10}

\definecolor{fluorescentpink}{rgb}{1.0, 0.08, 0.58}
\definecolor{D5Mr}{rgb}{0.62, 0.0, 1.0}
\definecolor{cUdg}{rgb}{0.3, 0.73, 0.09}
\definecolor{QKVd}{rgb}{0.03, 0.57, 0.82}
\definecolor{RE12}{rgb}{0.48, 0.25, 0.0}

\def\D5Mr#1{{\textcolor{D5Mr}{#1}}}

\def\RE12#1{{\textcolor{RE12}{#1}}}

\makeatletter
\long\def\@maketablecaption#1#2{\@tablecaptionsize
	\global \@minipagefalse
\hbox to \hsize{\parbox[t]{\hsize}{\centering #1 \\ #2}}}

\makeatother

\begin{document}
\allowdisplaybreaks
\begin{frontmatter}
	\title{Data-Driven Safety Certificates of Infinite Networks with Unknown Models and Interconnection Topologies\thanksref{footnoteinfo}}

\thanks[footnoteinfo]{Corresponding author:  Mahdieh Zaker (\texttt{m.zaker2@newcastle.ac.uk}).}

\author{Mahdieh Zaker}\ead{m.zaker2@newcastle.ac.uk},  
\author{Amy Nejati}\ead{amy.nejati@newcastle.ac.uk}, \author{Abolfazl Lavaei}\ead{abolfazl.lavaei@newcastle.ac.uk}  
\address{School of Computing, Newcastle University, United Kingdom}                              


\begin{abstract}   
	Infinite networks are complex interconnected systems comprising a countably infinite number of subsystems, for which no fixed upper bound on the number of participating subsystems is specified a priori since it may vary over time as agents join or leave (\emph{e.g.}, vehicles in traffic). In such scenarios, the presence of \emph{infinitely many} subsystems within the network renders the existing analysis frameworks tailored for finite networks inapplicable to infinite ones. This paper is concerned with offering a data-driven approach, within a compositional framework, for the safety certification of infinite networks with both \emph{unknown} mathematical models and \emph{unknown} interconnection topologies. Given the immense computational complexity stemming from the extensive dimension of infinite networks, our approach capitalizes on the joint dissipativity-type properties of subsystems, characterized by storage certificates. We introduce  innovative  compositional data-driven conditions to construct a barrier certificate for the infinite network leveraging storage certificates of its unknown subsystems derived from data, while offering  \emph{correctness guarantees} for network safety.  We demonstrate that our compositional data-driven reasoning eliminates the requirement for checking the traditional dissipativity condition, which typically mandates precise knowledge of the interconnection topology. We illustrate our data-driven results on two physical infinite networks with unknown models and interconnection topologies.
\end{abstract}

\end{frontmatter}

\section{Introduction}\label{Section: Introduction}
In today's world, various types of networks have become integral to daily life, spanning a wide range of applications from social to transportation networks. While such networks offer numerous benefits in various aspects of life, ensuring their safety certification requires meticulous attention due to the limited scalability of existing tools designed for large-scale systems. For instance, in connected vehicles, swarm robotics, smart cities, drone formations, and road traffic networks, subsystems (a.k.a. agents) can join or leave the network at any time. Hence, the sizes of these large networks are \emph{unknown} and potentially \emph{variable} over time~\citep{bamieh2002distributed, jovanovic2005ill,noroozi2021set,noroozi2022relaxed}. Consequently, treating these systems as finite networks would yield unrealistic models that fail to capture the genuine complexity of real-world scenarios.

To cope with this challenge, the notion of infinite networks has emerged, where finite yet exceedingly large networks are over-approximated by \emph{countably infinitely many} finite-dimensional subsystems~\citep{dashkovskiy2020stability}. This formulation provides a mathematical representation for networks whose exact finite size is unknown, potentially very large, and not bounded a priori. For instance, consider a traffic network composed of vehicles on the road: the total number of vehicles cannot be specified a priori, \emph{i.e.}, no fixed upper bound exists, and the network typically involves a large number of vehicles.

To illustrate why existing approaches designed for finite networks (\emph{i.e.,} large-scale networks with a known number of subsystems) cannot be applied to infinite networks, let us present a control problem example borrowed from~\citep{dashkovskiy2019stability}. In fact, this example demonstrates that while a finite cascade of input-to-state stable (ISS) subsystems remains ISS, this property does not generally hold for infinite networks. Consider the cascade system
\begin{align*}
\dot{x}_1=-x_1+u,\quad\dot{x}_{k+1}=-x_{k+1}+2x_k,\quad  k\in\mathbb{N}^+\!,
\end{align*}
for $x_i\in\mathbb{R},\, i\in\mathbb{N}^+$, where each scalar subsystem is ISS. However,  when considering an infinite number of subsystems with constant input $u=1$ and $x_i(0)=1$, the derivatives $\dot{x}_j,\, j\ge2$ become strictly positive for all times and $x_k(t)\to2^{k-1}$ as $t\to\infty,\,\forall k\in\mathbb{N}^+$. Consequently, $\lim_{t\to\infty}\Vert x(t)\Vert_\infty=\infty$, violating the ISS property of the overall infinite network~\citep{dashkovskiy2019stability}. This simple example demonstrates that methods developed for finite networks may fail when applied to infinite networks.

In the pursuit of providing formal safety certificates for complex dynamical systems, barrier certificates (BCs) have been introduced in relevant literature~\citep{prajna2004safety}.  Specifically, BCs are akin to Lyapunov functions, defined over the system's state space, and governed by a set of inequalities involving both the function and the system's flow or one-step transition. These certificates are primarily beneficial for establishing their initial level set from a set of initial states, while effectively separating an unsafe region from the system trajectories, thereby offering  a formal (probabilistic) guarantee over the system's safety~\citep{ames2019control,luo2020multi,clark2021control,borrmann2015control,nejati2024context,wooding2024protect}.

While BCs offer benefits for system safety certification, they do not scale well with system dimensions, proving ineffective when dealing with networks of infinite dimensions. Moreover, they typically require precise knowledge of system dynamics, rendering them inapplicable in scenarios involving unknown models.

\textbf{Key contribution.} These significant challenges have prompted us to develop a compositional data-driven approach for providing a safety certificate over an infinite network, consisting of a countably infinite number of unknown subsystems, interconnected via an unknown topology. Our methodology leverages the joint dissipativity-type properties of subsystems, which are described by storage certificates (StC), designed from data, to amalgamate them into a BC designed for the infinite network. Through our newly developed compositional data-driven conditions,  we design a BC based on StCs of individual subsystems, derived from data, while guaranteeing the safety of the unknown infinite network. Our data-driven approach significantly mitigates sample complexity by reducing it from being exponential in the \emph{network size} to being exponential in the effective dimension of each \emph{subsystem}. Additionally, our data-driven compositional condition obviates the necessity for the traditional dissipativity condition, which demands precise knowledge of the interconnection topology.

\textbf{Previous studies on infinite networks.} 
Several investigations have explored \emph{stability} analysis over infinite networks, including studies into controllability analysis~\citep{enyioha2014controllability}, development of small-gain reasoning for input-to-state stability~\citep{bao2018nonlinear, dashkovskiy2020stability,kawan2020lyapunov,kawan2022lyapunov}, decentralized fixed-time input-to-state stabilization of infinite networks for switched nonlinear systems~\citep{pavlichkov2023decentralized}, and adaptive decentralized stabilization~\citep{pavlichkov2023adaptive}. Despite the valuable insights provided by these studies, none of them addresses the challenges posed by infinite networks with unknown dynamics of subsystems and unknown interconnection topologies. In particular, prior studies \citep{dashkovskiy2019stability, dashkovskiy2020stability} establish the stability of infinite networks using small-gain theorems and ISS Lyapunov functions. However, these studies do not investigate the construction of the required Lyapunov functions and the corresponding gains, whose availability remains essential for applying the results. In contrast, our method guarantees the safety of infinite networks by constructing the StCs and their associated gains directly from data, without assuming their availability a priori.

Additionally, to the best of our knowledge, there has been no prior work (even model-based) on the analysis of infinite networks using \emph{dissipativity reasoning}, as previous studies have all relied on the ISS property of underlying subsystems using small-gain reasoning. In contrast, our work is the first to introduce a compositional data-driven approach for infinite networks by leveraging joint dissipativity properties of subsystems, serving as a rigorous mathematical framework for the analysis of interconnected networks.

\textbf{Previous data-driven studies.} In the data-driven literature, several studies have also addressed the challenge posed by dynamical systems with unknown mathematical models. Existing findings encompass safety verification of unknown complex systems~\citep{fan2019data}, stability certificates for unknown switched linear systems \citep{kenanian2019data}, and the construction of barrier certificates for unknown discrete- and continuous-time systems~\citep{nejati2023formal}. While these works yield remarkable results, their methods are tailored to \emph{monolithic} systems and do not scale to large-scale networks, particularly those with infinitely many subsystems. Applying these approaches to interconnected networks leads to computational complexity that typically grows exponentially with the network size, making them intractable for large or infinite networks (cf. the case studies by \cite{nejati2023formal}, which are limited to systems of at most two dimensions). In contrast, our proposed framework handles large-scale networks with a countably infinite number of subsystems while improving the computational complexity by restricting its exponential growth to the subsystem level.

While a limited number of studies focus on (compositional) data-driven approaches to address higher-dimensional systems (\eg, \citet{zhang2023compositional,lavaei2023data,noroozi2021data, lavaei2022formal,wang2023data,baggio2021data,lavaei2023compositional}), they exclusively deal with networks with a \emph{finite} number of subsystems, and their findings cannot be extended to networks with infinitely many subsystems. In contrast, our proposed approach guarantees the safety of the network without knowing the number of subsystems. Moreover, in~\citep{noroozi2021data,lavaei2023compositional}, the compositional framework requires explicit satisfaction of a dissipativity condition to extend subsystem-level results to the entire network, which necessitates knowledge of the interconnection topology. In contrast, our method relaxes this requirement: through the proposed data-driven compositional conditions (cf. Theorem~\ref{Thm1}), the dissipativity condition is satisfied implicitly, eliminating the need to know the interconnection structure. In addition, while the studies by \cite{lavaei2023compositional} and \cite{noroozi2021data} ensure the safety with a \emph{probabilistic} confidence using a random sampling approach, our method provides a safety certificate with a correctness guarantee, achieved via our deterministic sampling strategy.

The only existing work that investigates the safety of infinite networks using data is the recent study by~\cite{aminzadeh2024compositional}. Our proposed approach differs from~\citep{aminzadeh2024compositional} in some key aspects. First, while~\cite{aminzadeh2024compositional} derive safety certificates using ISS properties of subsystems, we focus on establishing safety certificates for infinite networks through the use of dissipativity properties of individual subsystems. This shift allows us to bypass the need for the lower bound of the radially unbounded condition of the ISS property on the local BCs (see~\citet[condition (5a)]{aminzadeh2024compositional}), leading to more relaxed results and fewer required assumptions. Another significant difference is that unlike~\citep{aminzadeh2024compositional} which assumes the interconnection topology is known, we consider a more realistic scenario where the topology is entirely unknown. In such cases, the LMI condition required for model-based dissipativity theory (see~\citet[Chapter 2]{2016Murat}) cannot be directly verified. Instead, we propose a more relaxed compositional condition that can be checked via the solution of our scenario optimization problem using available data.

\section{Problem Description}\label{Problem_Description}

\subsection{Notation}
Sets of real, non-negative and positive real numbers are represented by $\mathbb{R},\mathbb{R}^+_0$, and $\mathbb{R}^+$, respectively. We denote sets of non-negative and positive integers by $\mathbb{N} \coloneq \{0,1,2,\ldots\}$ and $\mathbb{N}^+\coloneq\{1,2,\ldots\}$, respectively. Given $N$ objects $z_i$, which may be scalars, vectors, or matrices, we use the notation $[z_1;\ldots;z_N]$ to denote their block stacking. When each $z_i\in\R^{n_i}$ is a vector, this notation coincides with the usual column-wise vector stacking and yields a vector of dimension $\sum_i n_i$. More generally, when the objects $z_i$ have different types or dimensions, $[z_1;\ldots;z_N]$ is understood, with a slight abuse of notation, as a stacked collection of its constituent entries rather than as a single Euclidean vector. A vector $x$ composed of a \emph{countably infinite number} of real vectors $x_i\in\R^{n_i}$ is represented by $x=(x_i)_{i\in \N^+}$. We denote the $p$-norm of a finite-dimensional vector $x_i=[x_{1_i};\dots;x_{n_i}]\in\mathbb{R}^{n_i}$ by $\Vert x_i\Vert_{p}$, which is defined by $\|x_i\|_p=(\sum_{j = 1}^{n_i}|x_{j_i}|^p)^{\sfrac{1}{p}}$ for $p\in[1,\infty)$. For scalar arguments, $|\cdot|$ denotes the absolute value. We specifically denote the \emph{Euclidean} norm of a finite-dimensional vector $x_i\in\R^{n_i}$ with $\|x_i\|$. The infinity norm is denoted by $\|x_i\|_\infty=\max_{j\in\{1,\dots,n_i\}}|x_{j_i}|$. We use the notation $\ell^p$, $p\in[1,\infty)$, to refer to the \emph{Banach space} consisting of all sequences $x=(x_i)_{i\in\N^+}$ of real vectors with finite $\ell^p$ norm $|x|_p<\infty$, where $\vert x\vert_p = (\sum_{i = 1}^{\infty} \Vert x_i\Vert_p ^p)^{\sfrac{1}{p}}$. The space $\ell^\infty$ is the Banach space consisting of all sequences $x=(x_i)_{i\in\N^+}$ of real vectors with finite norm $|x|_\infty= \sup_{i\in\N^+}\|x_i\|_\infty <\infty.$ Given sets $X_i, i\in \N^+$, their Cartesian product is denoted by $\prod_{i\in \N^+}X_i$. We represent an empty set by $\emptyset$. The transpose of a matrix $P$ is denoted by $P^\top$.

\subsection{Infinite Networks}

We first present discrete-time nonlinear \emph{subsystems} in the subsequent definition, through whose interconnection we construct an infinite network.

\begin{definition} \label{discrete}
	A discrete-time nonlinear subsystem (dt-NS) $\Psi_{i}, i \in \mathbb{N}^+$,  can be described as
	$$
		\Psi_{i}=(X_i,D_i,f_i),
	$$
	where $X_i\subseteq \mathbb R^{n_i}$ is the compact state set of the dt-NS, $D_i \subseteq \mathbb R^{p_i}$ is the compact internal input set of the dt-NS, and $f_i\!:X_i{\times} D_i\rightarrow X_i$ is the transition map characterizing the evolution of the dt-NS. Such a dt-NS can be represented as
	\begin{equation}\label{Eq_1a}
		\Psi_i\!:x_i(k+1) = f_i(x_i(k),d_i(k)),\quad 
		k\in \mathbb{N},
		\quad
	\end{equation}
	where $x_i\!:\mathbb{N} \rightarrow X_i$ and $d_i\!:\mathbb{N} \rightarrow D_i$ are the state and internal input signals, respectively. Although the maps $f_i, i \in \mathbb{N}^+$ are unknown in our data-driven setting, they are assumed to be Lipschitz continuous with respect to $(x_i,d_i)$ uniformly in $i$.
\end{definition}

Internal inputs $d_i$ capture the effect of neighboring subsystems in the interconnection topology and will subsequently be employed to interconnect and structure the infinite network (cf. the interconnection constraint \eqref{eq:unknown topology}). In the following definition, we define an infinite network comprising a \emph{countably infinite number} of dt-NSs.

\begin{definition} \label{network}
	Given dt-NSs $\Psi_{i}=(X_i,D_i,f_i), i \in \mathbb{N}^+$, an infinite network $\Psi$ is denoted by $\Psi = \mathcal{I}(\Psi_i)_{i \in \mathbb{N}^+}$, in which the interconnection of subsystems complies with the following constraint:
	\begin{subequations}\label{eq:unknown topology}
		\begin{align}
			d_i=\mathcal M_i\big((x_j)_{j\in\mathbf M_i}\big),\quad i\in\N^+\!,
		\end{align}
		with $\mathcal M_i : \prod_{ j\in \mathbf M_i}X_j \rightarrow  D_i$, specifying the subsystems' interconnection structure, where $\mathbf M_i\subset\N^+\setminus\{i\}$ is the unknown finite index set of subsystems affecting $\Psi_i$. The network topology is described by the operator $\mathcal M:\prod_{i\in\N^+}X_i\to \prod_{i\in\N^+}D_i$, characterized as
		\begin{align}\label{eq:topology net}
			\mathcal M(x)\coloneq \big(\mathcal M_i((x_j)_{j\in\mathbf M_i})\big)_{i\in\N^+}.
		\end{align}
	\end{subequations}
	Such an infinite network can be defined as $ \Psi = (X, f)$, and is specified by
	\begin{equation}\label{Eq_1a1}
		\Psi\!:x(k+1) = {f(x(k))}, \quad k\in\mathbb{N},
	\end{equation}
	where
	\begin{align}
		&\bullet X \!=\!\big\{x \!=\! (x_i)_{i \in \mathbb{N}^+}\!:\! x_i \!\in\! X_i, \vert x\vert_\infty  \!=\! \sup_{i\in\N^+}   \Vert x_i\Vert_\infty\!<\!\infty \big\};\label{eq:state set}\\
		&\bullet f(x)= (f_i(x_i,d_i))_{i \in \mathbb{N}^+},\text{ with }f\!:X\rightarrow X.\notag
	\end{align}
	 The sequence $x_{x_0}: \mathbb N \rightarrow X$ that satisfies~\eqref{Eq_1a1} determines the \textit{state trajectory} of $\Psi$ starting from an initial state $x_0$.
\end{definition}

We assume that the network transition map $f$ is well-defined on $X$, namely, $f(x)\in X$ for all $x\in X$. Hence, for every initial state $x_0\in X$, the recursion~\eqref{Eq_1a1} generates a unique trajectory $x_{x_0}:\mathbb N\to X$. Therefore, the infinite network $\Psi=\mathcal I(\Psi_i)_{i\in\mathbb N^+}$ is well-posed.

\begin{remark}
	In our framework, the interconnection topology in~\eqref{eq:unknown topology}, including both the maps $\mathcal M_i$ and the index sets $\mathbf M_i$, is unknown, which is the case in numerous real-life scenarios. In addition, the topology can be of any arbitrary form. More precisely, the topology is embedded in the data collection, meaning that for some topologies, the internal-input data becomes sparse (\eg, in a cascade topology where each subsystem is influenced only by the previous one) or dense (\eg, in a topology where each subsystem is influenced by numerous, yet finitely many, other subsystems in the network, as commonly encountered in traffic networks). We emphasize that while the interconnection topology $\mathcal M$, constructed by the maps $\mathcal M_i$ and the sets $\mathbf M_i$, is unknown, the dimension of each $d_i$ is known and finite, \ie, $d_i\in D_i$, with $D_i\subseteq \R^{p_i}$, where $p_i$ is known. Moreover, although the topology itself is unknown, the internal input signal \(d_i\) is assumed to be measurable during data collection.
\end{remark}
Since this work focuses on the safety verification of infinite networks, we formally define the notion of safety as follows.
\begin{definition}\label{def:safety}
		Consider a safety specification $\Lambda = (X_{0}, X_{a})$, with $X_{0}, X_{a} \subseteq X$ representing the initial and unsafe sets of the infinite network $\Psi$, defined similar to~\eqref{eq:state set}, where $X_{0} \cap X_{a} = \emptyset$. Then, the infinite network $\Psi$ is said to be safe over an infinite time horizon, denoted by $\Psi \vDash \Lambda$, if all trajectories starting from $X_{0}$ avoid $X_{a}$ at all times.
\end{definition}

\begin{remark}
		Condition $X_{0} \cap X_{a} = \emptyset$ ensures that the initial and unsafe sets do not intersect. If this condition is violated, it implies that the system is unsafe from the outset, and any safety analysis becomes meaningless.
\end{remark}
In the following section, we present the concept of storage and barrier certificates for individual subsystems and infinite networks, respectively. These concepts play a crucial role in guaranteeing the network safety in the sense of Definition~\ref{def:safety}.

\section{Storage and Barrier Certificates} \label{barrier}
We begin by presenting the formal definition of storage certificates for individual subsystems, affected by internal inputs.
\begin{definition} \label{def:csbc}
	Consider a dt-NS $\Psi_i= (X_i,D_i,f_i)$ as in Definition~\ref{discrete}. Let $X_{0_i}, X_{a_i} \subseteq X_i$ be initial and unsafe sets of the dt-NS, respectively. A function $\mathds B_i:X_i\rightarrow\mathbb{R}^+_0$ is called a storage certificate (StC) for $\Psi_i$ if there exist a symmetric block matrix $\mathcal S_i$, partitioned into $\mathcal S_i^{jj'}$, $j,j'\in\{1,2\}$, $\sigma_i, \phi_i ,c_i\in \R^+$, and $\lambda_i\in (0, 1)$, such that
	\begin{subequations}\label{eq:SBC all cons}
		\begin{align}
			\mathds B_i(x_i) &\leq \sigma_i,~~\qquad\!\forall x_i \in X_{0_i},\label{eq:subsys1}\\
			\mathds B_i(x_i) &\geq \phi_i, ~~\qquad\!\forall x_i \in X_{a_i},\label{eq:subsys3}
		\end{align}
		and $\forall x_i\in X_i, \;\forall d_i\in D_i,$
		\begin{align}
			\mathds B_i(f_i(x_i,d_i)) - \lambda_i\mathds B_i(x_i) &\leq \begin{bmatrix}
				d_i\\
				x_i
			\end{bmatrix}^{\!\top}\!\!
			\underbrace{\begin{bmatrix}
					\mathcal S_i^{11}&\mathcal S_i^{12}\\
					\mathcal S_i^{21}&\mathcal S_i^{22}
			\end{bmatrix}}_{\mathcal S_i}\!\begin{bmatrix}
				d_i\\
				x_i
			\end{bmatrix}\! + c_i,\label{eq:csbceq}
		\end{align}
	where $\mathcal S_i^{11}\in\mathbb R^{p_i\times p_i}$, $\mathcal S_i^{12}\in\mathbb R^{p_i\times n_i}$, $\mathcal S_i^{21} = \mathcal S_i^{{12}^\top}$, and $\mathcal S_i^{22}\in\mathbb R^{n_i\times n_i}$.
	\end{subequations}
\end{definition}

	Storage certificates $\mathds B_i$ encapsulate internal inputs $d_i$ resulting from the interplay between subsystems within the interconnection topology. This is elucidated through the inclusion of the quadratic term in the right-hand side of \eqref{eq:csbceq}. In dissipativity theory, this term represents the system's supply rate, as discussed by \cite{2016Murat}, initially employed to demonstrate the stability of an interconnected network comprising a finite number of subsystems.

We now present barrier certificates for infinite networks in the following definition, adapted from~\citep{akbarzadeh2025learning}.

\begin{definition} \label{def:cbc}
	Consider an infinite network $\Psi = (X, f)$, composed of a countably infinite number of dt-NSs $\Psi_{i}$, as in Definition~\ref{network}. A function $\mathds B:X \rightarrow \mathbb{R}_0^+$ is a barrier certificate (BC) for $\Psi$ if there exist $\sigma,\phi,c \in \R^+$, with $\phi > \sigma$, and $\lambda\in(0,1)$, such that
	\begin{subequations}\label{eq:BC all cons}
		\begin{align}
			&\mathds B(x) \leq \sigma,\quad\quad\quad\quad\quad\!\forall x \in X_{0},\label{eq:con1_BNetwork}\\
			&\mathds B(x) \geq \phi, \quad\quad\quad\quad\quad\!\forall x \in X_{a},\label{eq:con2_BNetwork}\\
			\mathds B(f(x)) - \lambda&\mathds B(x) \leq c, \quad\quad\quad\quad\quad\forall x\in X,\label{eq:cbceq}
		\end{align}
		where sets $X_0, X_a \subseteq X$ represent initial and unsafe sets of the infinite network, respectively, and $c$ satisfies
		\begin{align}\label{eq:inf time con}
			c\leq(1 -\lambda) \phi.
		\end{align}
	\end{subequations}
\end{definition}
The subsequent proposition, adapted from \citep{akbarzadeh2025learning}, assures that all state trajectories of the interconnected network refrain from entering an unsafe region within an infinite time horizon.
\begin{proposition}\label{theorem:safe}
	Given an infinite network $\Psi = (X, f)$ as in Definition~\ref{network}, let $\mathds{B}$ be a BC for $\Psi$ as in Definition~\ref{def:cbc}. Then, the network $\Psi$ is always safe within an infinite time horizon with respect to
	$\Lambda=(X_0,X_a)$, \ie, $x_{x_0}(k) \notin X_a$ for any $x_0 \in X_0$ and any $k \in \mathbb{N}$.
\end{proposition}
While Proposition~\ref{theorem:safe} ensures safety of infinite networks, designing   a BC generally  poses significant computational challenges, even when dealing with finite networks. That is why our compositional approach concentrates on subsystem levels, by initially constructing StCs and then under specific compositional conditions, deriving BC from StCs of individual subsystems. Nonetheless, it is evident that directly satisfying condition \eqref{eq:csbceq} for the StC construction is impractical due to the presence of \emph{unknown} models $f_i$. Considering these significant challenges, we introduce our data-driven scheme in the subsequent section.

\section{Data-Driven Construction of Storage Certificates}\label{Data-Driven}

In our data-driven scheme, we specify the structure of StC as
\begin{align}\label{Eq:3}
	\mathds{B}_i(\vartheta_i,x_i)=\sum_{j=1}^{l} {\vartheta}_i^j\varpi_i^j(x_i), 
\end{align}
with $\varpi_i^j(x_i)$ being user-defined (potentially nonlinear) basis functions of $x_i,$ and $\vartheta_i=[\vartheta_i^1;\dots;\vartheta_i^l]\in\R^l$ being \emph{unknown} decision variables to be designed. For instance, choosing $\varpi_i^j(x_i)$ as monomials in $x_i$ results in a polynomial $\mathds B_i(\vartheta_i, x_i)$. We assume that $\mathds B_i$ is continuously differentiable, which is required for establishing the main results presented in Theorem~\ref{Thm1}.
\begin{remark}
		The storage certificate $\mathds B_i(\vartheta_i, x_i)$ in our framework serves as a Lyapunov-like function, similar in nature to barrier certificates. As is well known in nonlinear control theory, such functions typically yield only sufficient conditions, and designing them can be potentially challenging. Consequently, when such a function cannot be found, no definitive conclusion about the system’s safety can be drawn. The choice of the basis functions $\varpi_i^j(x_i)$ is crucial and since the system dynamics are unknown, one can leverage physical intuition from the black-box model to inform the choice of basis functions. For instance, in the case of a jet engine, one might reasonably assume that the system exhibits polynomial dynamics due to its underlying physical properties, whereas for a quadcopter, trigonometric functions may be more appropriate due to its rotational motion.
\end{remark}
To facilitate understanding as we develop the technical concepts in this work, we introduce our first case study as a running example.

\textbf{Running example.}
	 Consider a building temperature network comprising a countably infinite number of rooms, characterized by unknown mathematical models and interconnection topology. Each room within this network is equipped with a cooler~\citep{meyer}, with the state space as $X_i = [10, 13]$, the initial set is $X_{0_i}=[10, 11]$, and the unsafe set is $X_{a_i}=[12, 13]$. We set the structure of our StC as $\mathds B_i(\vartheta_i,x_i) = \vartheta_i^1 x_i^4  + \vartheta_i^2 x_i^2 + \vartheta_i^3$, for all $i \in \mathbb{N}^+$. In other words, $\varpi_i^1(x_i) = x_i^4,\, \varpi_i^2(x_i) = x_i^2, \, \varpi_i^3(x_i)=x_i^0=1$. $\hfill\square$

We first reformulate the essential conditions of StC in Definition~\ref{def:csbc} into a robust convex program (RCP), for a fixed $\lambda_i$, as follows:
\begin{mini!}|s|[2]
	{[\Omega_i;\eta_{R_i};\beta_{R_i}]}{\eta_{R_i} + \beta_{R_i},}{\label{RCP}}{}\notag
	\addConstraint{{\forall x_i\!\in\! X_{i_{\hphantom{0}}}\!\!:-\mathds B_i(\vartheta_i,x_i)} }{ \leq \eta_{R_i},\label{RCP1}}
	\addConstraint{\forall x_i\!\in\! X_{0_i}\!\!:\hphantom{-}\mathds B_i(\vartheta_i,x_i) - \sigma_i}{ \leq \eta_{R_i},\label{RCP2}}
	\addConstraint{\forall x_i\!\in\! X_{a_i}\!\!:  - \mathds B_i(\vartheta_i,x_i) + \phi_i\leq \eta_{R_i},\label{RCP2-1}}
	\addConstraint{\forall (x_i,d_i) \in X_i\times D_i\!:}{}\notag
	\addConstraint{\mathds B_i(\vartheta_i,f_i(x_i,d_i)) - {\lambda_i}\mathds B_i(\vartheta_i,x_i)}{}\notag
	\addConstraint{~-\begin{bmatrix}
			d_i\\
			x_i
		\end{bmatrix}^\top\!
		\begin{bmatrix}
			\mathcal S_i^{11}&\mathcal S_i^{12}\\
			\mathcal S_i^{21}&\mathcal S_i^{22}
		\end{bmatrix}\begin{bmatrix}
			d_i\\
			x_i
	\end{bmatrix}{-c_i}}{\leq \eta_{R_i},\label{RCP2-2}}
	\addConstraint{\begin{bmatrix}
			d_i\\
			x_i
		\end{bmatrix}^\top\!\!
		\begin{bmatrix}
			\mathcal S_i^{11}&\mathcal S_i^{12}\\
			\mathcal S_i^{21}&\mathcal S_i^{22}
		\end{bmatrix}\!\begin{bmatrix}
			d_i\\
			x_i
	\end{bmatrix}}{\leq \beta_{R_i},\label{RCP3}}
	\addConstraint{\Omega_i = [\sigma_i;\phi_i;{\vartheta}_{i}^1;\dots;\vartheta_{i}^l;{c_i};\mathcal S_i^{11};\mathcal S_i^{12};\mathcal S_i^{22}],}{}\notag
	\addConstraint{\sigma_i,\phi_i,{c_i}\!\in\!{\R^+}\!,{\vartheta}_{i}^j,\eta_{R_i},\beta_{R_i} \!\in \!\mathbb R,{\lambda_i\!\in\!(0,1)}.}\notag
\end{mini!}
To ensure that the RCP is well-posed, we assume that all decision variables in $\Omega_i$ are constrained to lie within compact sets. We denote the optimal value of RCP as $\eta^*_{R_i} + \beta^*_{R_i}$. If $\eta^*_{R_i} \leq 0$, it is clear that a solution to the RCP ensures the fulfillment of the conditions outlined in Definition~\ref{def:csbc}, thus establishing $\mathds B_i$ as an StC for the dt-NS in~\eqref{Eq_1a}.  It is evident that the RCP exhibits convexity with respect to decision variables; thanks to the choice of $\mathds B_i$ in \eqref{Eq:3}. Of note is, since $\lambda_i$ is fixed a priori, it is not considered as a decision variable and the RCP remains a convex problem.

To solve the proposed RCP in~\eqref{RCP}, one may experience two crucial impediments. First, since the state and internal input sets $X_i\subseteq\R^{n_i}$ and $D_i\subseteq\R^{p_i}$ are not finite, the proposed RCP consists of infinitely many constraints. Additionally, solving the RCP directly is intractable as satisfying condition~\eqref{RCP2-2} necessitates the knowledge of model $f_i$, which is unknown in our setting. To overcome these critical challenges, we aim to construct an StC via developing a data-driven approach without directly solving the RCP in~\eqref{RCP}. In doing so, we first gather pairs $\big((x_i^z, d_i^z), f_i(x_i^z, d_i^z)\big),\; z\in\{1,\dots,\mathcal N_i\}$ of two consecutive samples from unknown subsystems' trajectories, with $\mathcal N_i$ being the required number of sample pairs. The sample set covers $X_{0_i}$, $X_{a_i}$, and $X_i\times D_i$ in a way compatible with the relevant constraints. We then compute the maximum distance between any pair $(x_i, d_i)\in X_i\times D_i $ and the set of all sample pairs as follows:
\begin{align}
	\theta_{i} = \max_{(x_i,d_i)}\min_z&\Vert (x_i,d_i) - (x^z_i,d^z_i)\Vert, \quad\forall i\in\N^+\!\!.\label{eq:epsilon}
\end{align}
While $\theta_i$ in~\eqref{eq:epsilon} is computed over $X_i\times D_i$, one can restrict attention to $X_{0_i}$ and $X_{a_i}$ and compute $\theta_i^0$ and $\theta_i^a$, respectively, in a manner similar to~\eqref{eq:epsilon}. By considering $(x^z_i,d^z_i)\in X_i\times D_i, z\in\{1,\dots,\mathcal N_i\}$, we now propose the following scenario convex program (SCP), for a fixed $\lambda_i$, purely based on collected data:
\begin{mini!}|s|[2]
	{[\Omega_i;\eta_{\mathcal N_i};\beta_{\mathcal N_i}]}{\eta_{\mathcal N_i} + \beta_{\mathcal N_i},}{\label{SCP}}{}\notag
	\addConstraint{\forall z\in\{1,\dots,\mathcal N_i\}}{}\notag
	\addConstraint{{\forall x_i^z\in X_{i_{\hphantom{0}}}\!\!:-\mathds B_i(\vartheta_i,x^z_i)} }{\leq \eta_{\mathcal N_i},\label{SCP1}}
	\addConstraint{\forall x_i^z\in X_{0_i}\!\!:\hphantom{-}\mathds B_i(\vartheta_i,x^z_i) - \sigma_i}{ \leq \eta_{\mathcal N_i},\label{SCP2}}
	\addConstraint{\forall x_i^z\in X_{a_i}\!\!:- \mathds B_i(\vartheta_i,x^z_i) + \phi_i}{\leq \eta_{\mathcal N_i},\label{SCP2-1}}
	\addConstraint{\forall(x^z_i,d^z_i)\in X_i\times D_i:}{}\notag
	\addConstraint{\mathds B_i(\vartheta_i,f_i(x^z_i,d^z_i)) -{\lambda_i} \mathds B_i(\vartheta_i,x^z_i)}{}\notag
	\addConstraint{~-\begin{bmatrix}
			d^z_i\\
			x^z_i
		\end{bmatrix}^\top\!
		\begin{bmatrix}
			\mathcal S_i^{11}&\mathcal S_i^{12}\\
			\mathcal S_i^{21}&\mathcal S_i^{22}
		\end{bmatrix}\begin{bmatrix}
			d^z_i\\
			x^z_i
	\end{bmatrix}{-c_i}}{\leq \eta_{\mathcal N_i},\label{SCP2-2}}
	\addConstraint{\begin{bmatrix}
			d^z_i\\
			x^z_i
		\end{bmatrix}^\top\!
		\begin{bmatrix}
			\mathcal S_i^{11}&\mathcal S_i^{12}\\
			\mathcal S_i^{21}&\mathcal S_i^{22}
		\end{bmatrix}\begin{bmatrix}
			d^z_i\\
			x^z_i
	\end{bmatrix}}{\leq \beta_{\mathcal N_i},\label{SCP3}}
	\addConstraint{\Omega_i = [\sigma_i;\phi_i;{\vartheta}_{i}^1;\dots;\vartheta_{i}^l;{c_i};\mathcal S_i^{11};\mathcal S_i^{12};\mathcal S_i^{22}],}{}\notag
	\addConstraint{\sigma_i,\phi_i,{c_i}\!\in\!{\R^+}\!,{\vartheta}_{i}^j,\eta_{\mathcal N_i},\beta_{\mathcal N_i} \!\in \!\mathbb R,{\lambda_i\!\in\!(0,1)}.}\notag
\end{mini!}
It is evident that  the problem of unknown $f_i$ in the proposed SCP~\eqref{SCP} is resolved, as the second data point within the pair $\big((x_i^z, d_i^z), f_i(x_i^z, d_i^z)\big),\; z\in\{1,\dots,\mathcal N_i\}$, precisely encapsulates unknown subsystems' dynamics. We denote the optimal value of SCP as $\eta_{\mathcal N_i}^* + \beta_{\mathcal N_i}^*$, and construct the StC from the obtained decision variables $\vartheta_i^*$ as $ \mathds B_i(\vartheta_i^*,\cdot)$.

\begin{remark}
	The magnitude of the supply rate plays a key role in the compositional conditions we propose based on data (cf. Theorem~\ref{Thm1}). This underscores the primary rationale behind incorporating an additional condition in~\eqref{RCP3} and~\eqref{SCP3}, where the objective of the optimization problems involves the minimization of both $\eta_i$ and $\beta_i$. In particular, since the left-hand side of~\eqref{SCP3} appears explicitly in our compositional condition in \eqref{Con2}, including this constraint during the design phase serves to limit the magnitude of this term and facilitates the satisfaction of the compositional condition.
\end{remark}

\textbf{Running example (cont.).}
	We collect $\mathcal N_i = 12000$ sample pairs $\big((x_i^z, d_i^z), f_i(x_i^z, d_i^z)\big)$ and compute ${\theta_i^0}={\theta_i^a}=\theta_i={0.0071}$ given the collected data. {We fix $\lambda_i=0.1$}, and by solving SCP~\eqref{SCP} for all $i \in \mathbb{N}^+$, we obtain all decision variables as follows:
\begin{align}\label{sim-res}
	&\mathds B_i(\vartheta_i^*,x_i) = \underbrace{{0.0153}}_{{\vartheta_i^{1*}}}x_i^4  \underbrace{-0.7}_{{\vartheta_i^{2*}}}x_i^2 \underbrace{-0.7}_{{\vartheta_i^{3*}}},\\\notag
	&\mathcal{S}_i^{11^*} \!\!= 0.01,~ \mathcal{S}_i^{12^*} \!\!= \mathcal{S}_i^{21^*} \!\!= 0,~  \mathcal{S}_i^{22^*} \!\!= -0.1,~\sigma_i^* = 150, \\\notag
	& \phi_i^*\!=\!200, \;{c_i^*\!=\!116.2013}, \; {\beta_{\mathcal N_i}^* \!=\! -9.9526},\; {\eta_{\mathcal N_i}^* \!=\! -11.3927}.
\end{align}
Here, all subsystems are assumed to be identical for the sake of simplicity. $\hfill\square$

In the following section, by introducing our compositional data-driven conditions, we construct a BC for an infinite network with an unknown interconnection topology, via its StCs of individual subsystems, constructed through solving the proposed SCP, thus guaranteeing the safe operation of the infinite network.

\section{Safety Certificate across Infinite Networks}\label{Guarantee_RCP}
In our context, we assume that each unknown $f_i$ is Lipschitz continuous with respect to $(x_i,d_i)$,  a well-defined assumption in practical scenarios. Given that the StC $\mathds B_i$ is continuously differentiable and we perform our analysis on the \emph{compact domain} $X_i\times D_i$, one can easily come to the conclusion that $\mathds B_i(\vartheta_i^*,f_i(x_i,d_i)) - {\lambda_i}\mathds B_i(\vartheta_i^*,x_i)$ is Lipschitz continuous with respect to $(x_i,d_i)$ as well, with a Lipschitz constant $\mathcal{L}^2_{i}$. Analogously, with a similar justification, one can readily show that $\mathds B_i(\vartheta_i^*,x_i)$ is also Lipschitz continuous with respect to $x_i$, with a Lipschitz constant $\mathcal{L}^1_{i}$. In a later stage, we provide an algorithm to derive Lipschitz constants $\mathcal{L}^1_{i}$ and $\mathcal{L}^2_{i}$ only from collected data. 

Before introducing the main result of the work, we present the following assumption required to ensure that the BC and its level sets are well-defined.

\begin{assumption}\label{asmp:mu}
	Let $\mathds B_i(\vartheta_i^*,\cdot)$, $\sigma_i^*$, $\phi_i^*$, and $c_i^*$,
	$i\in\mathbb N^+$, be obtained from the solution of the SCP in~\eqref{SCP}.
	We assume that there exist
	$\mu=(\mu_i)_{i\in\mathbb N^+}\in\ell^1$, with
	$\mu_i\in\mathbb R^+$ for all $i\in\mathbb N^+$, and constants
	$\bar{\mathds B},\bar\sigma,\bar\phi,\bar c\in\mathbb R^+$, such that
	$$
	\sum_{i\in\mathbb N^+}\mu_i<\infty,~ \mathds B_i(\vartheta_i^*,x_i)\leq \bar{\mathds B},~\sigma_i^*\leq \bar\sigma, ~\phi_i^*\leq \bar\phi,~ c_i^*\leq \bar c,
	$$
	for all $x_i\in X_i$ and all $i\in\mathbb N^+$\!.
\end{assumption}
The requirement $\mu\in\ell^1$ in Assumption~\ref{asmp:mu} is mild; for instance, choosing $\mu_i=2^{-i}$ gives $\sum_{i\in\mathbb N^+}\mu_i=\sum_{i=1}^{\infty}2^{-i}=1<\infty.$ This assumption ensures that the candidate BC in~\eqref{eq:cbc-network} is well-defined
and finite. In addition, we note that for each fixed $i$, compactness of $X_i$ and continuity of $\mathds B_i(\vartheta_i^*,\cdot)$ imply boundedness of $\mathds B_i(\vartheta_i^*,\cdot)$ over $X_i$. Assumption~\ref{asmp:mu} requires this upper boundedness to hold uniformly with respect to the subsystem index $i$. This condition is automatically satisfied for homogeneous networks and, more generally, for networks consisting of finitely many subsystem classes.

We now offer our compositional data-driven conditions in the following theorem as the main finding of our work.
\begin{theorem}\label{Thm1}
	Consider an infinite network $\Psi = \mathcal{I}(\Psi_i)_{i \in \mathbb{N}^+}$, composed of a countably infinite number of unknown subsystems~$\Psi_i$, with an unknown interconnection topology. Let SCP in~\eqref{SCP} be solved with optimal values $\eta_{\mathcal N_i}^*+\beta_{\mathcal N_i}^*$, and solutions $\Omega_i^* = [\sigma_i^*;\phi_i^*;{\vartheta}^{1*}_i;\dots;\vartheta^{l*}_i;{c_i^*};\mathcal S_i^{11^*};\mathcal S_i^{12^*};\mathcal S_i^{22^*}]$. Under Assumption~\ref{asmp:mu}, if
	\begin{subequations}\label{eq:comp data-driven}
		\begin{align}
				& \sum_{i\in\mathbb N^+}{\mu_i}\sigma^*_i <  \sum_{i\in\mathbb N^+}{\mu_i}\phi^*_i, \label{Con0}\\
				&{\sum_{i\in \N^+}\mu_i c_i^*\leq\big(1 - \sup_{i\in\N^+}\{\lambda_i\}\big)\sum_{i\in \N^+}\mu_i\phi_i^*,}\label{Con0-1}\\
				&{\mu_i\big(\eta_{\mathcal N_i}^* \!+\! \mathcal{L}^1_{i} \bar\theta_{i} \big) \leq 0,\quad\forall i\in\N^+,}\label{Con1}\\
				&{\sum_{i\in \mathbb N^+}\mu_i\big(\eta_{\mathcal N_i}^* \!+\! \beta_{\mathcal N_i}^* \!+\! \mathcal{L}^2_{i} \theta_{i} \big) \leq 0,}\label{Con2}
			\end{align}
	\end{subequations}
	where $\bar\theta_i = \max\{\theta_i^0,\theta_i^a,\theta_i\}$, then  $\mathds B$ in the form of
	\begin{equation}\label{eq:cbc-network}
		{\mathds B(\vartheta,x)\coloneq\sum_{i\in \mathbb N^+}\mu_i\mathds B_i(\vartheta_i^*,x_i)<\infty,}
	\end{equation}
	is a BC for the infinite network $\Psi = \mathcal{I}(\Psi_i)_{i \in \mathbb{N}^+}$, satisfying conditions~\eqref{eq:con1_BNetwork}--\eqref{eq:inf time con}, with
	\begin{subequations}\label{eq:cbc params}
		\begin{align}
			&{\sigma\coloneq\sum_{i\in \N^+}\mu_i\sigma_i^* < \infty,\quad \phi \coloneq \sum_{i\in \N^+} \mu_i\phi_i ^*< \infty,}\label{eq:net level}\\
			&{c\coloneq\sum_{i\in \N^+}\mu_i c_i^* < \infty,\quad\lambda\coloneq\sup_{i\in\N^+}\{\lambda_i\}<1.\label{eq:lam_c_net}}
		\end{align}
	\end{subequations}
	Consequently, the infinite network $\Psi$ is safe within an infinite time horizon, according to Proposition~\ref{theorem:safe}, with a correctness guarantee.
\end{theorem}
\noindent\textbf{Proof.} We first show that each StC is nonnegative over its state set. For any $x_i\in X_i$, let $\tilde z\coloneq \arg \underset{z}{\min} \Vert x_i - x^z_i \Vert$, with $x_i^z\in X_i$. We add and subtract the term $\mathds B_i(\vartheta_i^*,x_i^{\tilde z})$ to $-\mathds B_i(\vartheta_i^*,x_i)$. Using the Lipschitz continuity of $\mathds B_i(\vartheta_i^*,\cdot)$, one obtains
\begin{align*}
	-\mathds B_i(\vartheta_i^*,x_i) &= -\mathds B_i(\vartheta_i^*,x_i) + \mathds B_i(\vartheta_i^*,x_i^{\tilde z}) - \mathds B_i(\vartheta_i^*,x_i^{\tilde z})\\
	&\leq -\mathds B_i(\vartheta_i^*,x_i^{\tilde z}) + \mathcal L_i^1\min_z\|x_i-x_i^z\|.
\end{align*}
Since $\underset{z}{\min} \Vert x_i - x^z_i\Vert \leq \underset{z}{\min} \Vert (x_i, d_i) - (x^z_i, d^z_i) \Vert$, according to~\eqref{eq:epsilon} and~\eqref{SCP1},
we have
\begin{align*}
	-\mathds B_i(\vartheta_i^*,x_i)
	&\!\leq \!\!-\mathds B_i(\vartheta_i^*,x_i^{\tilde z}) \!+\! \mathcal L_i^1\min_z\|(x_i, d_i) \!-\! (x^z_i, d^z_i)\|\\
	&\!\leq\!\!-\mathds B_i(\vartheta_i^*,x_i^{\tilde z}) \!+\! \mathcal L_i^1\!\max_{(x_i,d_i)}\!\min_z\|(x_i, d_i)\! -\! (x^z_i, d^z_i)\|\\
	&\!=\!-\mathds B_i(\vartheta_i^*,x_i^{\tilde z}) + \mathcal L_i^1\theta_i  \leq\eta_{\mathcal N_i}^*+ \mathcal L_i^1\theta_i.
\end{align*}
According to~\eqref{Con1}, and since $\mu_i>0$ and $\bar\theta_i = \max\{\theta_i^0,\theta_i^a,\theta_i\}$, it follows that
$$
\mathds B_i(\vartheta_i^*,x_i)\geq0,\qquad
\forall x_i\in X_i,\ \forall i\in\mathbb N^+.
$$
Now, we show that the BC defined in~\eqref{eq:cbc-network} is well-defined
and finite. By Assumption~\ref{asmp:mu}, and by the nonnegativity established
above, it follows that, for all $x\in X$,
$$
0\leq \mathds B(\vartheta,x) = \sum_{i\in \mathbb N^+}\mu_i\mathds B_i(\vartheta_i^*,x_i)\leq \bar{\mathds B}\sum_{i\in \mathbb N^+}\mu_i<\infty.
$$
Similarly, according to Assumption~\ref{asmp:mu}, one can deduce that $0<\sigma=\sum_{i\in \N^+}\mu_i\sigma_i^* < \infty$, $0<\phi= \sum_{i\in \N^+} \mu_i\phi_i ^*< \infty$, and $0<c=\sum_{i\in \N^+}\mu_i c_i^* < \infty$.

We continue with demonstrating that under condition~\eqref{Con1}, $\mathds B$ in the form of~\eqref{eq:cbc-network} satisfies condition~\eqref{eq:con1_BNetwork} over the initial set, \emph{i.e.,}
$$
\mathds B(\vartheta, x) - \sigma\leq 0,\quad\forall x \in X_{0}.
$$
By considering $\mathds B(\vartheta, x)$ based on StCs of individual subsystems as in \eqref{eq:cbc-network}, we have
$$
\mathds B(\vartheta,x) - \sigma = \sum_{i\in \mathbb N^+}{\mu_i}  \mathds B_i(\vartheta_i^*,x_i) -\sigma,
$$
with $x_i\in X_{0_i}$. Let $\tilde z\coloneq \arg \underset{z}{\min} \Vert x_i - x^z_i \Vert$, with $x_i^z\in X_{0_i}$.  By incorporating the terms $\sum_{i\in \mathbb N^+}{\mu_i}\mathds B_i(\vartheta_i^*,x^{\tilde z}_i)$ through addition and subtraction, and considering $\sigma\coloneq \sum_{i\in \mathbb N^+}  \mu_i\sigma^*_i<\infty$ as in~\eqref{eq:net level}, one has
\begin{align*}
	\mathds B(\vartheta,x) - \sigma = \sum_{i\in \mathbb N^+} &{\mu_i}\big(\mathds B_i(\vartheta_i^*,x_i) \!-\! \sigma^*_i \!+\! \mathds B_i(\vartheta_i^*,x^{\tilde z}_i)\\
	& ~~- \mathds B_i(\vartheta_i^*,x^{\tilde z}_i)\big).
\end{align*}
In view of the fact that $\mathds B_i(\vartheta_i^*,x_i)$ is Lipschitz continuous with respect to $x_i$ with a Lipschitz constant $\mathcal{L}^1_{i}$, one has
$$
\mathds B(\vartheta,x)- \sigma \leq \!\sum_{i\in \mathbb N^+} \!{\mu_i}\big(\mathcal{L}^1_{i}\min_z\Vert x_i - x^z_i \Vert\!+\! \mathds B_i(\vartheta_i^*,x^{\tilde z}_i) - \sigma^*_i\big).
$$
Since $\underset{z}{\min} \Vert x_i - x^z_i\Vert \leq \underset{z}{\min} \Vert (x_i, d_i) - (x^z_i, d^z_i) \Vert$, {and according to the definition of $\theta_i^0$ similar to~\eqref{eq:epsilon},} for any $x_i, x_i^z\in X_{0_i}$, one can obtain
\begin{align*}
	\mathds B(\vartheta,x)- \sigma &\leq \sum_{i\in \mathbb N^+} {\mu_i}\big(\mathcal{L}^1_{i}\min_z\Vert (x_i, d_i) - (x^z_i, d^z_i)\Vert\\
	&\hspace{1.2cm}+ \mathds B_i(\vartheta_i^*,x^{\tilde z}_i) \!-\! \sigma_i^*\big)\\
	&\leq \sum_{i\in \mathbb N^+}\!{\mu_i}\big(\mathcal{L}^1_{i}\! \max_{(x_i,d_i)}\!\min_z\!\Vert (x_i, d_i) - (x^z_i, d^z_i)\Vert\\
	&\hspace{1.2cm}+\mathds B_i(\vartheta_i^*,x^{\tilde z}_i) \!-\! \sigma_i^*\big)\\
	& = \sum_{i\in \mathbb N^+}{\mu_i}\big(\mathcal{L}_{i}^1 {\theta_i^0} + \mathds B_i(\vartheta_i^*,x^{\tilde z}_i) \!-\! \sigma_i^*\big).
\end{align*}
According to condition~\eqref{SCP2} of the SCP, and since $\bar\theta_i = \max\{\theta_i^0,\theta_i^a,\theta_i\}$, we have
$$
\mathds B(\vartheta,x)- \sigma \leq \sum_{i\in \mathbb N^+}{\mu_i}\big(\mathcal{L}^1_{i}{\bar\theta_{i}}+\eta_{\mathcal N_i}^*\big).
$$
Given the proposed condition in~\eqref{Con1}, one can conclude that
$$
\mathds B(\vartheta, x) - \sigma\leq 0,\quad\forall x \in X_{0},
$$
with  $\sigma= \sum_{i\in \mathbb N^+} {\mu_i} \sigma^*_i<\infty$. Following similar steps, one can show that according to condition~\eqref{Con1}, $\mathds B$ in the form of~\eqref{eq:cbc-network} satisfies condition~\eqref{eq:con2_BNetwork} with $\phi= \sum_{i\in \mathbb N^+}  {\mu_i}\phi_i^*<\infty$. Note that $\sigma < \phi$ according to condition~\eqref{Con0}.

Now, we proceed with showing that under condition~\eqref{Con2}, $\mathds B$ in the form of~\eqref{eq:cbc-network} satisfies condition~\eqref{eq:cbceq} as well, for the whole range of the state set, \emph{i.e.,}
$$
\mathds B(\vartheta,f(x)) - {\lambda}\mathds B(\vartheta,x) \leq {c}, \quad \quad \forall x\in X.
$$
In accordance with the proposed BC in~\eqref{eq:cbc-network} and the definition of $\lambda$ in~\eqref{eq:lam_c_net}, one has
\begin{align*}
	&\mathds B(\vartheta,f(x)) - {\lambda}\mathds B(\vartheta,x) \\
	&=\sum_{i\in \mathbb N^+}{\mu_i}\big(\mathds B_i(\vartheta_i^*,f_i(x_i,d_i)) - {\lambda} \mathds B_i(\vartheta_i^*,x_i)\big)\\
	&{=\sum_{i\in \mathbb N^+}\mu_i\big(\mathds B_i(\vartheta_i^*,f_i(x_i,d_i)) - \sup_{i\in\N^+}\{\lambda_i\} \mathds B_i(\vartheta_i^*,x_i)\big)}\\
	&{\leq\sum_{i\in \mathbb N^+}\mu_i\big(\mathds B_i(\vartheta_i^*,f_i(x_i,d_i)) - \lambda_i \mathds B_i(\vartheta_i^*,x_i)\big),}
\end{align*}
where $x_i\in X_i$. Let $\tilde z\coloneq \arg \underset{z}{\min} \Vert (x_i,d_i) - (x^z_i,d^z_i) \Vert$, with $x_i^z\in X_i$.  By incorporating the terms $\sum_{i\in \mathbb N^+}{\mu_i}\big(\mathds B_i(\vartheta_i^*,$ $f_i(x^{\tilde z}_i,d^{\tilde z}_i)) - {\lambda_i}\mathds B_i(\vartheta_i^*,x^{\tilde z}_i)\big)$ through addition and subtraction, one attains
\begin{align*}
	&\mathds B(\vartheta,f(x)) \!-\! {\lambda}\mathds B(\vartheta,x) \leq\\
	& \sum_{i\in \mathbb N^+}\!\!{\mu_i}\big(\mathds B_i(\vartheta_i^*,f_i(x_i,d_i)\!) \!-\! {\lambda_i} \mathds B_i(\vartheta_i^*,x_i) \!-\!\mathds B_i(\vartheta_i^*,f_i(x^{\tilde z}_i,d^{\tilde z}_i)\!) \\
	&\hphantom{\sum_{i\in \mathbb N^+}\!\!\mu_i\big(}\!+\!{\lambda_i}\mathds B_i(\vartheta_i^*,x^{\tilde z}_i)\!+\mathds B_i(\vartheta_i^*,f_i(x^{\tilde z}_i,d^{\tilde z}_i)\!)\!-\! {\lambda_i}\mathds B_i(\vartheta_i^*,x^{\tilde z}_i)\big).
\end{align*}
Since $\mathds B_i(\vartheta_i^*,f_i(x_i,d_i)) -{\lambda_i} \mathds B_i(\vartheta_i^*,x_i)$ is Lipschitz continuous with respect to $(x_i,d_i)$ with a Lipschitz constant $\mathcal{L}_{i}^2$, and as per~\eqref{eq:epsilon}, we have
\begin{align*}
	\mathds B(\vartheta&,f(x)) - {\lambda}\mathds B(\vartheta,x) \\
	&\leq \sum_{i\in \mathbb N^+}{\mu_i}\big(\mathcal{L}_{i}^2 \, \min_z\Vert (x_i,d_i) - (x^z_i,d^z_i)\Vert\\
	&\hphantom{\leq \sum_{i\in \mathbb N^+}\mu_i\big(}+ \mathds B_i(\vartheta_i^*,f_i(x^{\tilde z}_i,d^{\tilde z}_i))-{\lambda_i} \mathds B_i(\vartheta_i^*,x^{\tilde z}_i)\big)\\
	&\leq \sum_{i\in \mathbb N^+}{\mu_i}\big(\mathcal{L}_{i}^2 \, \max_{(x_i,d_i)}\min_z\Vert (x_i,d_i) - (x^{z}_i,d^{z}_i)\Vert\\
	&\hphantom{\leq \sum_{i\in \mathbb N^+}\mu_i\big(}+ \mathds B_i(\vartheta_i^*,f_i(x^{\tilde z}_i,d^{\tilde z}_i)) - {\lambda_i}\mathds B_i(\vartheta_i^*,x^{\tilde z}_i)\big)\\
	& = \sum_{i\in \mathbb N^+}{\mu_i}\big(\mathcal{L}_{i}^2 \theta_i + \mathds B_i(\vartheta_i^*,f_i(x^{\tilde z}_i,d^{\tilde z}_i)) - {\lambda_i}\mathds B_i(\vartheta_i^*,x^{\tilde z}_i)\big).
\end{align*}
According to conditions~\eqref{SCP2-2} and \eqref{SCP3} of SCP, one has
\begin{align*}
	\mathds B(\vartheta,f(x)) - {\lambda}\mathds B(\vartheta,x)&\leq \!\!\sum_{i\in \mathbb N^+}\!\!{\mu_i}\big(\mathcal{L}_{i}^2 \theta_{i} + \eta_{\mathcal N_i}^* + \beta_{\mathcal N_i}^*{+c_i^*}\big)\\
	&{= \!\!\!\sum_{i\in \mathbb N^+}\!\!\mu_i\big(\mathcal{L}_{i}^2 \theta_{i} \!+\! \eta_{\mathcal N_i}^* \!+\! \beta_{\mathcal N_i}^*\big)\!+\!\!\!\sum_{i\in \mathbb N^+}\!\!\mu_ic_i^*.}
\end{align*}
Considering $c\coloneq \sum_{i\in \mathbb N^+}  \mu_i c_i^*<\infty$ as in~\eqref{eq:lam_c_net}, one gets
$$
{\mathds B(\vartheta,f(x)) - \lambda\mathds B(\vartheta,x)-c\leq\sum_{i\in \mathbb N^+}\!\!\mu_i\big(\mathcal{L}_{i}^2 \theta_{i} \!+\! \eta_{\mathcal N_i}^* \!+\! \beta_{\mathcal N_i}^*\big).}
$$
Given the proposed condition in~\eqref{Con2}, it can be deduced that
$$
\mathds B(\vartheta,f(x)) - {\lambda}\mathds B(\vartheta,x) \leq {c}, \quad \quad \forall x\in X\!,
$$
satisfying condition~\eqref{eq:cbceq}. Finally, as per~\eqref{eq:cbc params}, condition~\eqref{Con0-1} ensures the satisfaction of condition~\eqref{eq:inf time con}. Then, one can conclude that $\mathds B(\vartheta, x)$ in the form of~\eqref{eq:cbc-network} is a BC for the infinite network $\Psi$,  with a \emph{correctness guarantee},  thus ensuring the network safety, which completes the proof.$\hfill\blacksquare$

\begin{remark}
	Condition~\eqref{Con2} involves an infinite summation and should be understood as a well-defined inequality. A simple sufficient requirement is that the term $\eta_{\mathcal N_i}^*+\beta_{\mathcal N_i}^*+\mathcal L_i^2\theta_i$ is uniformly bounded from below with respect to $i$. Since $\mu\in\ell^1$ and $\mu_i>0$, this ensures that, if condition~\eqref{Con2} holds, then $-\infty<\sum_{i\in\mathbb N^+}\mu_i \big(\eta_{\mathcal N_i}^*+\beta_{\mathcal N_i}^*+\mathcal L_i^2\theta_i\big)\leq 0$. This requirement is automatically satisfied for homogeneous networks and, more generally, for networks with finitely many subsystem classes.
\end{remark}

The data-driven conditions in Theorem~\ref{Thm1} are designed to ensure that the BC defined in~\eqref{eq:cbc-network} satisfies the safety conditions of Definition~\ref{def:cbc}. Specifically, condition~\eqref{Con0} ensures that the network-level constants associated with the initial and unsafe sets are strictly separated, namely $\sigma<\phi$, as required by Definition~\ref{def:cbc}. It is noteworthy that the proposed compositional condition does not require $\sigma_i^*<\phi_i^*$ for every individual subsystem. Condition~\eqref{Con1} guarantees the fulfillment of conditions \eqref{eq:con1_BNetwork} and \eqref{eq:con2_BNetwork} across the network through satisfying conditions \eqref{eq:subsys1} and~\eqref{eq:subsys3} per subsystem by considering the BC as defined in~\eqref{eq:cbc-network}, while also ensuring the nonnegativity of the BC. Specifically, the term $\eta_{\mathcal{N}_i}^* + \mathcal{L}^1_i {\bar\theta_i}$ quantifies the conservatism level required for robustness to \emph{unseen} data. Condition~\eqref{Con2} accounts for the unknown system dynamics and supply rate to ensure the satisfaction of~\eqref{eq:cbceq} through \eqref{eq:csbceq}. To formally quantify the required negative margin to account for unseen data, we use the Lipschitz constant $\mathcal{L}_i^2$ and the optimal bound $\beta_{\mathcal{N}_i}^*$ such that the summation of $\mu_i\big(\eta_{\mathcal N_i}^* \!+\! \beta_{\mathcal N_i}^* \!+\! \mathcal{L}^2_{i} \theta_{i} \big)$ in \eqref{Con2} across the network remains non-positive. This implies that some subsystems may have a positive value of $\mu_i\big(\eta_{\mathcal N_i}^* \!+\! \beta_{\mathcal N_i}^* \!+\! \mathcal{L}^2_{i} \theta_{i} \big)$, provided that the other subsystems compensate for the adverse effect of this positivity so that the overall summation across the network remains non-positive. Collectively, these conditions guarantee that the BC defined in~\eqref{eq:cbc-network} satisfies the safety requirements of Definition~\ref{def:cbc}, even under limited data.
	
It is worth noting that the proposed conditions in Theorem~\ref{Thm1} are merely sufficient, which is well known in the analysis of \emph{nonlinear} systems using methods based on Lyapunov or barrier functions. Hence, BCs cannot certify that the system is unsafe when the specified conditions are violated. In such cases, a common remedy is to increase the degree of the barrier certificate in the case of polynomial BC when no feasible solution is found for a given degree. This increases the number of decision variables, making the certificate more expressive and thereby improving the likelihood of successful design. However, this also comes with increased computational cost, as an expected trade-off.
\begin{algorithm}[t!]
	\caption{Estimation of Lipschitz constants $\mathcal{L}_{i}^1$ and $\mathcal{L}_i^2$ using data}\label{alg:1}
	\begin{center}
		\begin{algorithmic}[1]
			\REQUIRE 
			StC $\mathds B_i(\vartheta_i^*,\cdot)$
			\STATE 
			Choose $\bar{\alpha}, \tilde{\alpha} \in \mathbb{N}^+$ and $\delta\in \mathbb{R}^+$\label{step:1}
			\FOR{$j \in  \{1,\dots,\tilde{\alpha}\}$}
			\FOR{$z \in \{1, \dots, \bar{\alpha}\}$}
			\STATE
			Collect  sampled pairs $((x_i^z,d_i^z), (\hat x_i^z,\hat d_i^z))$ such that $\Vert  (x_i^z,d_i^z) - (\hat x_i^z,\hat d_i^z) \Vert \leq \delta$\label{step:4}
			\STATE
			Compute the slope $$\rho_i^z = \dfrac{\Vert  \Gamma(x_i^z,d_i^z) - \Gamma(\hat x_i^z,\hat d_i^z)  \Vert}{\Vert   (x_i^z,d_i^z ) - (\hat x_i^z,\hat d_i^z) \Vert},$$ with $\Gamma(x_i^z,d_i^z) = \mathds B_i(\vartheta_i^*, f_i(x_i^z, d_i^z)) -{\lambda_i} \mathds B_i(\vartheta_i^*, x_i^z),$ ($\Gamma(\hat x_i^z,\hat d_i^z)$ is obtained similarly)
			\ENDFOR
			\STATE
			Derive the maximum slope as $\mathcal R_j^i = \max \{\rho_i^1,\dots,\rho_i^{\bar{\alpha}}\}$
			\ENDFOR
			\STATE
			Utilizing the \textit{Reverse Weibull distribution} on $\mathcal R_1^i, \dots, \mathcal R_{\tilde{\alpha}}^i$~\citep{wood1996estimation}, which results in location, scale, and shape parameters, opt for the \textit{location parameter} as an estimation of $\mathcal{L}^2_{i}$\label{step:10}
			\STATE
			Repeat Steps \ref{step:1}--\ref{step:10} with $\Gamma(x_i^z) = \mathds B_i(\vartheta_i^*, x_i^z)$ to estimate $\mathcal{L}^1_{i}$
			\ENSURE $\mathcal{L}^1_{i},\mathcal{L}^2_{i}$ 
		\end{algorithmic}
	\end{center}
\end{algorithm}
\begin{remark}
	As shown in~\citep{2016Murat}, for networks with certain interconnection topologies and supply rates, dissipativity-based compositional conditions recover classical small-gain-type conditions. For instance, in a two-scalar-subsystem network with interconnection matrix $\mathcal M$ (with a slight abuse of notation compared to the nonlinear map~\eqref{eq:topology net}) and supply rate matrices $\mathcal{S}_i $, respectively, as
	$$
	\mathcal{M} = \begin{bmatrix} 0 & -1 \\ 1 & 0 \end{bmatrix}\!\!,\;\text{and}\;\;
	\mathcal{S}_i = \begin{bmatrix} \gamma_{i(3-i)}^2 & 0 \\ 0 & -1 \end{bmatrix}\!\!,\quad i\in\{1,2\},
	$$
	the dissipativity-based stability condition, with appropriate positive scaling multipliers in the aggregate supply rate, recovers the classical small-gain criterion $\gamma_{12} \gamma_{21} \leq 1$. Moreover, for large-scale networks with skew-symmetric interconnection matrices ($\mathcal{M}^\top = -\mathcal{M}$), passivity-based dissipativity conditions can naturally exploit the interconnection structure, regardless of the number of subsystems, providing scalability not generally guaranteed by the small-gain approach. We note that our framework enables data-driven compositional safety guarantees without requiring knowledge of the interconnection matrix. This flexibility allows our approach to accommodate a wider variety of system architectures and uncertain interconnections.
\end{remark}
To verify our proposed data-driven compositional conditions~\eqref{Con1}--\eqref{Con2}, one requires the Lipschitz constants $\mathcal{L}_i^1$ and $\mathcal{L}_i^2$. We leverage the foundational result of \citep{wood1996estimation} and provide Algorithm~\ref{alg:1} to estimate these Lipschitz constants for individual subsystems $\Psi_i$ using a finite number of samples. The convergence of the estimated values $\mathcal{L}_i^1$ and $\mathcal{L}_i^2$ to their exact values follows from the following lemma~\citep{wood1996estimation}.
\begin{lemma}\label{lemma: lip}
	The estimated values $\mathcal{L}_i^1$ and $\mathcal{L}_i^2$ converge to their actual values if and only if $\delta$  approaches zero while $\bar{\alpha}$ and $\tilde{\alpha}$ tend to infinity.
\end{lemma}
\begin{remark}
	Algorithm~\ref{alg:1} mainly relies on pairwise evaluations of data samples within two nested loops. Accordingly, the dominant computational cost arises from performing $\bar{\alpha} \tilde{\alpha}$ slope computations and updates per subsystem, resulting in a time complexity of $\mathcal{O}(\bar{\alpha} \tilde{\alpha})$, assuming that the cost of evaluating $\mathds B_i$ and fitting the Reverse-Weibull distribution is either constant or lower order. The repetition of the procedure to estimate $\mathcal{L}_i^1$ only doubles this cost and therefore does not alter the overall order of complexity. Regarding memory usage, provided that the samples and slopes are processed sequentially, the algorithm stores $\bar{\alpha}$ sample pairs and at most $\max\{\bar{\alpha}, \tilde{\alpha}\}$ auxiliary scalars, leading to a memory complexity of $\mathcal{O}(\max\{\bar{\alpha}, \tilde{\alpha}\})$. It is worth noting that the neighborhood condition in Step~\ref{step:4} used to enforce bounded pairwise distances does not incur additional computational overhead when the data are actively collected, as it can be satisfied by collecting the first arbitrary samples $(x_i^z, d_i^z)\in X_i \times D_i$ and subsequently choosing the corresponding $(\hat{x}_i^z, \hat{d}_i^z)$ to be sufficiently close, such that their Euclidean distance does not exceed $\delta$.
\end{remark}
We now provide Algorithm~\ref{alg:2} to summarize our compositional data-driven scheme for constructing a BC for an infinite network based on StCs of its individual subsystems. It is noteworthy that, for homogeneous networks, the loops in Algorithm~\ref{alg:2} are executed only once for the representative subsystem, whereas for networks with finitely many subsystem classes, they are executed once per class.

\begin{algorithm}[t!]
	\caption{Compositional data-driven construction of BC}\label{alg:2}
	\begin{center}
		\begin{algorithmic}[1]
			\REQUIRE Structure of StC $\mathds B_i$
			\FOR{$z\in \{1, \dots, \mathcal N_i\}$,}\label{step 1}
			\STATE
			Collect sample pairs $\big(\!(x_i^z, \!d_i^z), \!f_i(x_i^z, \!d_i^z)\!\big)$ from each unknown subsystem's trajectory
			\ENDFOR
			\FOR{each unknown dt-NS $\Psi_i, i \in \mathbb{N}^+$,}
			\STATE
			Solve SCP \eqref{SCP} and attain optimal value $\eta_{\mathcal N_i}^*+\beta_{\mathcal N_i}^*$, and solution $\Omega_i^* = [\sigma_i^*;\phi_i^*;{\vartheta}^{1*}_i;\dots;\vartheta^{l*}_i;{c_i^*};\mathcal S_i^{11^*};\mathcal S_i^{12^*};\mathcal S_i^{22^*}]$
			\STATE
			Estimate Lipschitz constants $\mathcal{L}_i^1$ and $\mathcal{L}_i^2$ through Algorithm~\ref{alg:1}
			\STATE
			Compute  $\theta_i$ as in \eqref{eq:epsilon}, together with $\theta_i^0$ and $\theta_i^a$ over their associated state sets
			\STATE
			Obtain $\bar\theta_i = \max\{\theta_i^0, \theta_i^a,\theta_i\}$
			\STATE
			Compute $\eta_{\mathcal N_i}^* \!+\! \mathcal{L}_i^1 {\bar\theta_i}$ and $\eta_{\mathcal N_i}^* \!+\! \beta_{\mathcal N_i}^* \!+\! \mathcal{L}_i^2 \theta_i$
			\ENDFOR
			\IF{compositional conditions \eqref{Con0}--\eqref{Con2} hold}
			\STATE
			$\mathds B(\vartheta, x) \coloneq \sum_{i\in \N^+} {\mu_i}\mathds B_i(\vartheta_i^*, x_i)$ is a BC for the network $\Psi$
			\ELSE
			\STATE
			Return to Step \ref{step 1} and collect more sample pairs (\ie, smaller $\theta_i$ or $\bar\theta_i$), potentially facilitating the satisfaction of compositional conditions
			\ENDIF
			\ENSURE BC $\mathds B$, network's safety
		\end{algorithmic}
	\end{center}
\end{algorithm}
\begin{figure}[t!]
	\centering
	\subfloat[Satisfaction of conditions~\eqref{eq:subsys1}--\eqref{eq:subsys3}\label{fig:barrier}]{
		\resizebox{0.75\linewidth}{!}{
			\begin{tikzpicture}
				\begin{axis}[
					colorbar,colormap/bluered,
					xmin=10, xmax = 13,
					xtick={10, 11, 12, 13},
					xlabel= Temperature,
					ylabel=$\mathds B_i(x_i)$,
					samples=200,
					grid=both, major grid style={black!50,dashed},
					point meta rel=per plot]
					\addplot[mesh,ultra thick, domain=10:13,colormap/bluered] {0.0153*x^4-0.7*x^2-0.7};
					\addplot[dashed,green!70!blue,ultra thick, domain=10:13] {150};
					\addplot[dashed,red!70,ultra thick, domain=10:13] {200};
					
					\draw[thick, draw=green!30, fill=green!30,fill opacity=0.2] (axis cs:10, 0) rectangle (axis cs:11, 400);
					
					\draw[thick, draw=red!30, fill=red!30,fill opacity=0.2] (axis cs:12, 0) rectangle (axis cs:13, 400);
					
					\node at (axis cs:10.25, 315) [anchor=west] {$X_{0_i}$};
					\node at (axis cs:12.25, 315) [anchor=west] {$X_{a_i}$};
					
					\node at (axis cs:11.35, 135) [anchor=west] {\textcolor{green!70!blue}{$\sigma_i$}};
					\node at (axis cs:11.35, 215) [anchor=west] {\textcolor{red!70}{$\phi_i$}};
					
				\end{axis}
			\end{tikzpicture}
	}}
	\\
	\subfloat[Satisfaction of condition~\eqref{eq:csbceq}\label{fig:heat}]{
		\includegraphics[width=0.65\linewidth]{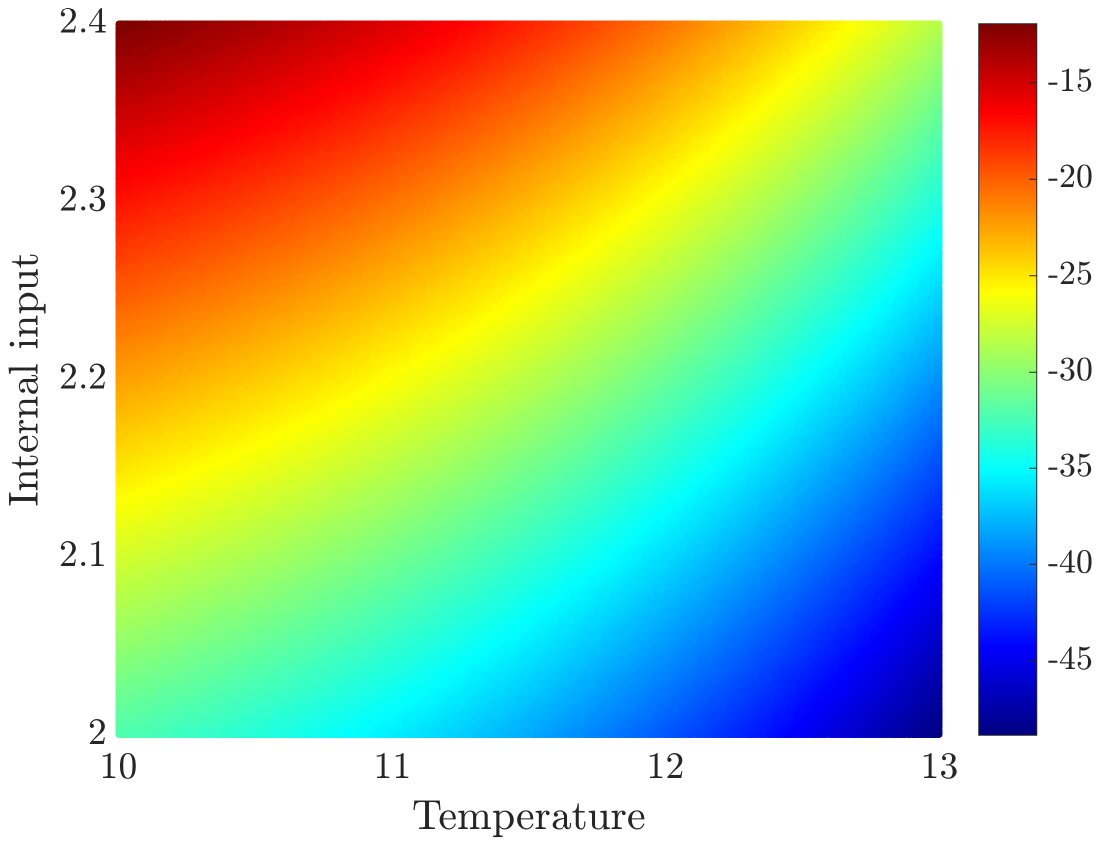}}
	\\
	\subfloat[Trajectories of arbitrary rooms\label{fig:room traj}]{
		\includegraphics[width=0.63\linewidth]{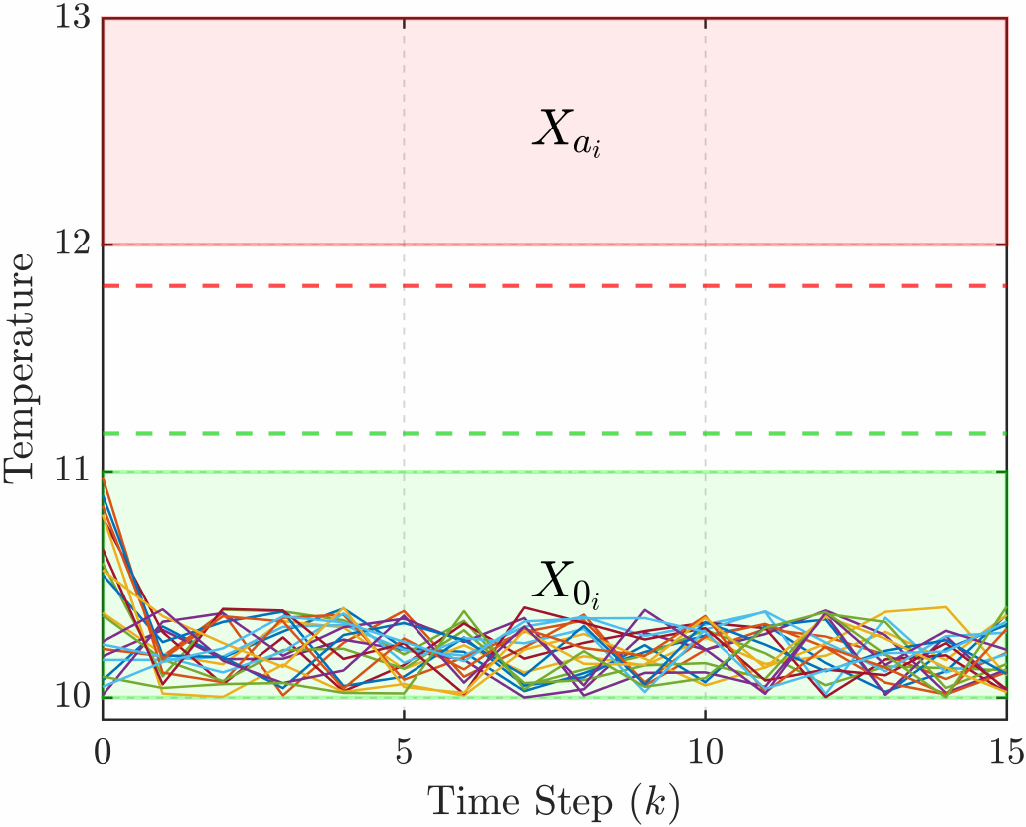}}
	\caption{{\bf Room temperature network:} Fig. (a) depicts $\mathds B_i(x_i)$ with its corresponding colorbar, with initial and unsafe level sets $\mathds B_i(x_i) = \sigma_i$ (\sampleline{dashed, gdash, thick}) and $\mathds B_i(x_i) = \phi_i$ (\sampleline{dashed, red, thick}), ensuring the fulfillment of conditions~\eqref{eq:subsys1} and~\eqref{eq:subsys3}. Fig. (b) demonstrates the satisfaction of  condition~\eqref{eq:csbceq} as the heatmap is negative for the whole range of $X_i$ and $D_i$ when shifting the supply rate and $c_i$ to the left-hand side of~\eqref{eq:csbceq}. Fig. (c) depicts that no trajectory originating from the initial set $X_{0_i}$ (\legendsquare{green!8}) enters the unsafe set $X_{a_i}$ (\legendsquare{red!8}). \label{fig:room-barrier}}
\end{figure}

\textbf{Running example (cont.).}
	We choose $\mu_i = 2^{-i}$ complying with Assumption~\ref{asmp:mu}, which yields $\sum_{i\in\mathbb N^+}\mu_i=\sum_{i=1}^{\infty}2^{-i}=1$. Since the subsystems are identical, we have $\sigma = \sum_{i\in \N^+}\mu_i\sigma_i^* = \sigma_i^*\sum_{i\in \N^+}\mu_i = 150$, and $\phi = \sum_{i\in \N^+}\mu_i \phi_i^*= \phi_i^*\sum_{i\in \N^+}\mu_i =200$, demonstrating the satisfaction of condition~\eqref{Con0}. Additionally, we have $c = \sum_{i\in \N^+}\mu_i c_i^*= c_i^*\sum_{i\in \N^+}\mu_i =116.2013$ and $\lambda = \sup_{i\in\N^+}\{\lambda_i\} = 0.1$, fulfilling condition~\eqref{Con0-1}. According to Algorithm~\ref{alg:1}, we now  compute\footnote{The Lipschitz constants are assumed to be accurately estimated and available, without any confidence levels involved.} $\mathcal{L}_{i}^1 = {116.2564}$ and $\mathcal{L}_i^2 = {55.5127}$. Since $\eta_{\mathcal N_i}^* +\mathcal{L}^1_{i}{\bar\theta_i}={-10.5673} \leq 0$ and $\eta_{\mathcal N_i}^*+\beta_{\mathcal N_i}^*+\mathcal{L}^2_{i}\theta_i={-20.9512} \leq 0$, for all $i \in \mathbb{N}^+$, according to Theorem~\ref{Thm1}, $\mathds B(\vartheta,x)=\sum_{i \in \mathbb{N}^+}{\mu_i}\mathds B_i(\vartheta_i^*,x_i),$ with $\mathds B_i(\vartheta_i^*,x_i)$ as in~\eqref{sim-res}, is a BC for the unknown infinite network with unknown topology. $\hfill\square$
	
\section{Case Studies}\label{compu}
In this section, we verify the effectiveness of our data-driven findings through their application to two physical infinite networks: (i) building temperature (running example throughout the paper) and (ii) vehicle networks. Both cases involve large-scale networks with a \emph{countably infinite} number of subsystems with unknown mathematical models and unknown interconnection topologies.

To report the runtime, we decompose the procedure of Algorithm~\ref{alg:1} into three main components: (i) sample pair collection and slope computation $\rho_i^z$ within the \emph{inner} loop of size $\bar{\alpha}$ (with runtime $T_{\text{Alg.1}, 1}$), (ii) derivation of the maximum slopes $\mathcal{R}_j^i$ within the \emph{outer} loop of size $\tilde{\alpha}$ (with $T_{\text{Alg.1}, 2}$), and (iii) estimation of the Lipschitz constant via fitting a Reverse-Weibull distribution over $\mathcal{R}_1^i, \dots, \mathcal{R}_{\tilde{\alpha}}^i$ (with $T_{\text{Alg.1}, 3}$). Similarly, Algorithm~\ref{alg:2} is divided into two primary phases: the first step involves data collection and the construction of data-driven constraints (with runtime $T_{\text{Alg.2}, 1}$), and the second phase involves solving the resulting SCP using the MOSEK\footnote{To use MOSEK, one should first obtain a license from \url{https://www.mosek.com}.} solver (with $T_{\text{Alg.2}, 2}$). Then, we report the runtime values for each phase of these algorithms along with the total runtime per subsystem for each case study. The reported results are obtained via a MacBook Pro (Apple M2 Pro chip with $16$ GB memory). 

\subsection{Building Temperature Network}
The obtained results for this case study are already presented throughout the paper as a running example. The runtime values per subsystem are $T_{\text{Alg.1}, 1} \approx 0.0042$, $T_{\text{Alg.1}, 2} \approx 0.0422$, and $T_{\text{Alg.1}, 3} \approx 0.0034$, where $T_{\text{Alg.1},2}$ already includes $T_{\text{Alg.1},1}$.  Hence, the total runtime is $T_{\text{Alg.1}} \approx T_{\text{Alg.1},2}+T_{\text{Alg.1},3}\approx 0.0456$ seconds for Algorithm~\ref{alg:1}. Furthermore, the runtime for the first phase of Algorithm~\ref{alg:2} is $T_{\text{Alg.2}, 1} \approx {0.0024}$ and the solver latency is $T_{\text{Alg.2}, 2} \approx {0.0264}$, with a total runtime  $T_{\text{Alg.2}} \approx {0.0288}$ for each subsystem.
 The fulfillment of conditions \eqref{eq:subsys1}--\eqref{eq:csbceq} {and the trajectories of some arbitrary subsystems} can be seen in Fig. \ref{fig:room-barrier}.

\subsection{Vehicle Network}
We consider a vehicle network comprising a countably infinite number of vehicles with unknown mathematical models and unknown interconnection topology.

Each vehicle has two states, \ie, $x_i = [x_{1_i};x_{2_i}]$, with the state space $X_i = [0.8, 1.5]\times[0.8, 2]$, the initial set $X_{0_i} = [0.8, 1]\times[0.8, 1]$ and the unsafe set $X_{a_i} = [0.8, 1.5]\times[1.5, 2]$. To verify the network's safety via our proposed approach, we follow the proposed steps in Algorithm~\ref{alg:2}.

\begin{figure}[t!]
		\centering
	\subfloat[Phase portrait of the trajectories and satisfaction of conditions~\eqref{eq:subsys1}--\eqref{eq:subsys3}\label{fig:barrier2}]{
		\includegraphics[width=0.65\linewidth]{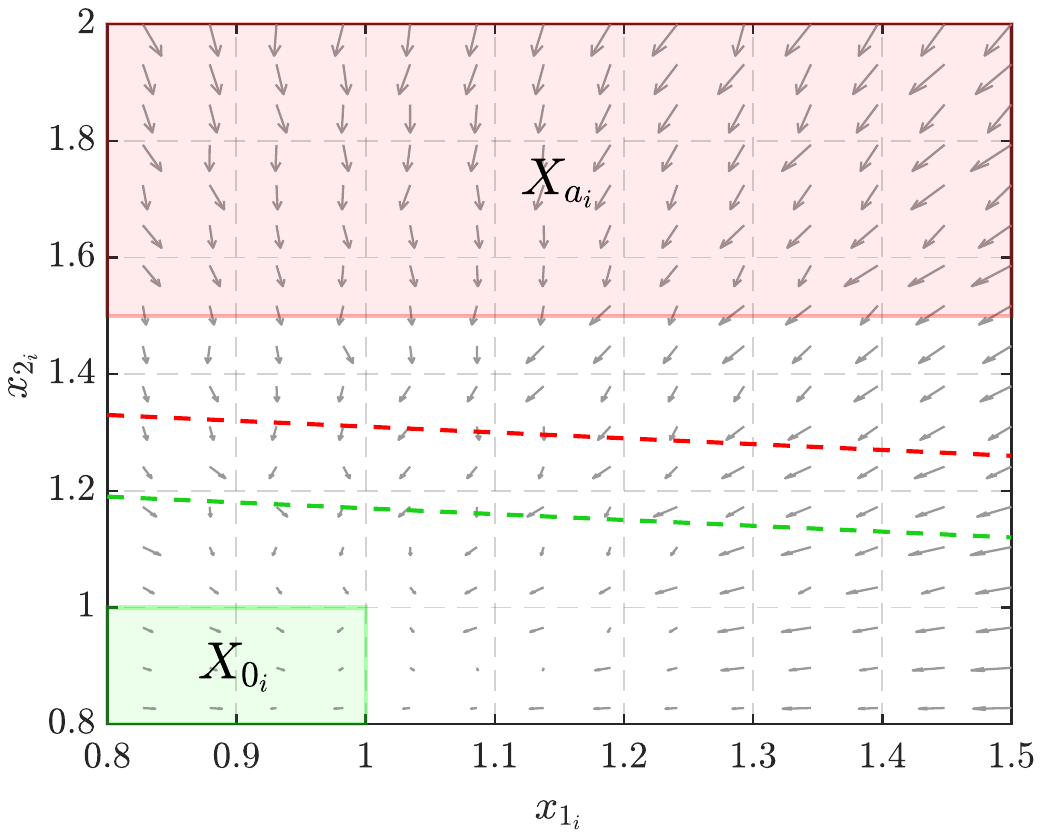}}
	\\
	\subfloat[Satisfaction of condition~\eqref{eq:csbceq}\label{fig:heat2}]{
		\includegraphics[width=0.7\linewidth]{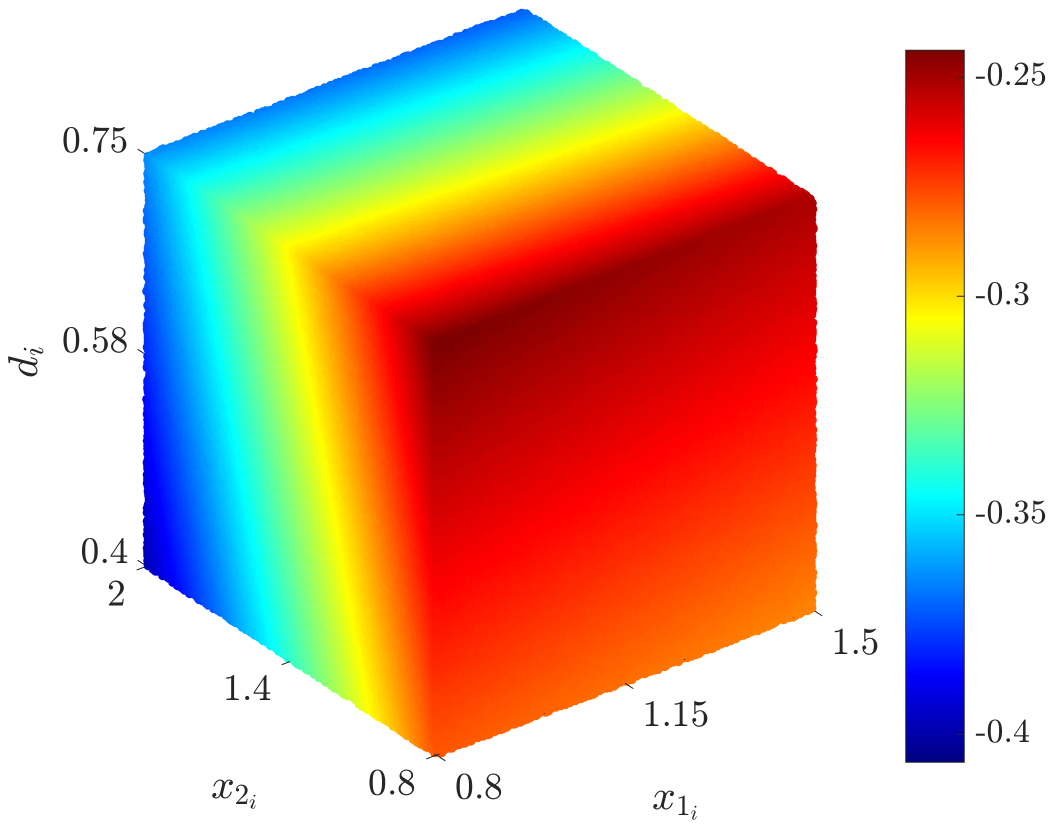}}
	\caption{{\bf Vehicle platooning network:} Fig. (a) depicts that no trajectory (\vectorarrow{thick, gray}) originating from the initial set $X_{0_i}$ (\legendsquare{green!8}) enters the unsafe set $X_{a_i}$ (\legendsquare{red!8}). As can be seen, the initial and unsafe level sets $\mathds B_i(x_i) = \sigma_i$ (\sampleline{dashed, gdash, thick}) and $\mathds B_i(x_i) = \phi_i$ (\sampleline{dashed, red, thick}) satisfy conditions~\eqref{eq:subsys1} and~\eqref{eq:subsys3}. Fig. (b) illustrates the satisfaction of  condition~\eqref{eq:csbceq}.\label{fig:platoon-res}}
\end{figure}

\begin{figure}[h!]
	\centering
	\resizebox{0.75\linewidth}{!}{
		\begin{tikzpicture}
			\node[inner sep=0pt, anchor=south west] (fig) 
			at (0,0) {\includegraphics[width=0.7\textwidth]{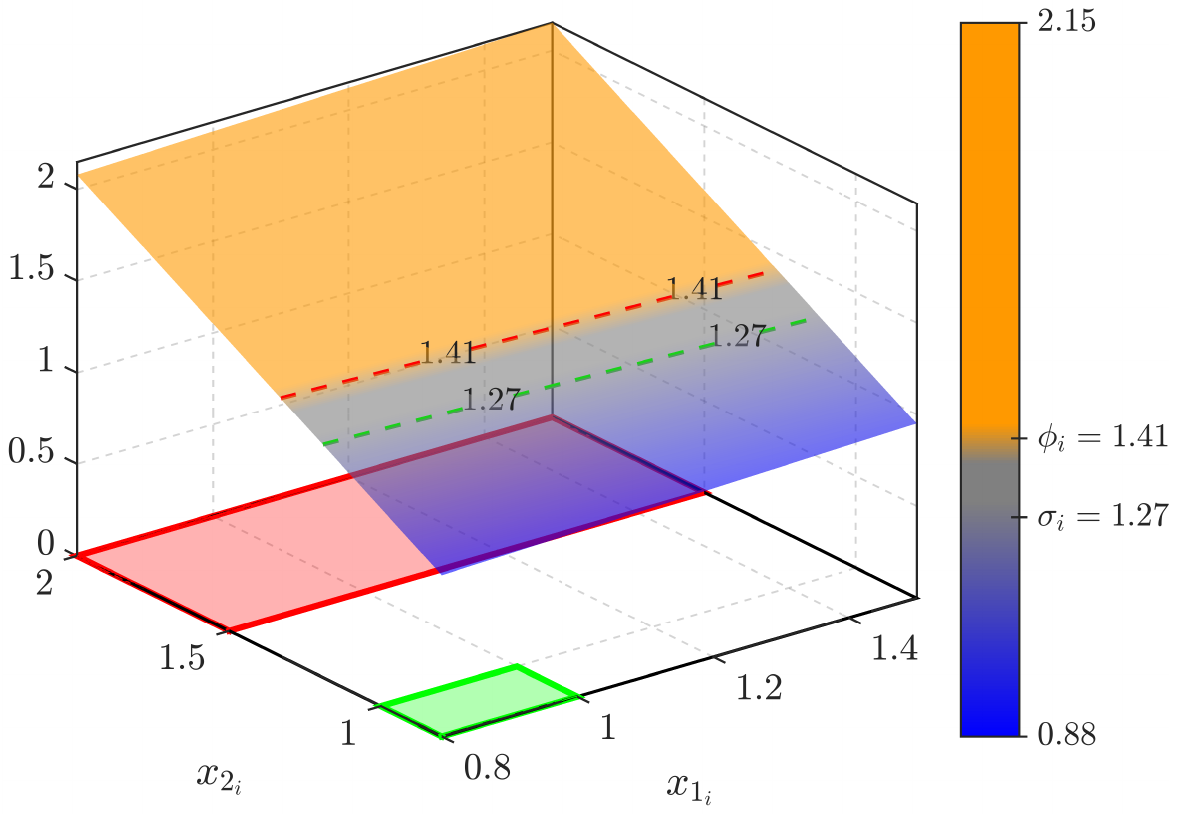}};

			\begin{scope}[x={(fig.south east)}, y={(fig.north west)}]
				\node[
				font=\fontsize{12pt}{12pt}\selectfont,
				rotate=90,
				anchor=center
				] at (-0.02,0.56) {$\mathds{B}_i(x_i)$};
			\end{scope}
		\end{tikzpicture}
	}
	\caption{Demonstration of the StC surface over the state space $X_i$ (\legendrect{very thick, black}). As can be seen, the value of StC is smaller than the contour $\mathds B_i(x_i) = \sigma_i$ (\sampleline{dashed, gdash, very thick}) over the initial set $X_{0_i}$ (\legendsquare{green!30}) and is greater than contour $\mathds B_i(x_i) = \phi_i$ (\sampleline{dashed, red, very thick}) over the unsafe set $X_{a_i}$ (\legendsquare{red!30}), satisfying conditions~\eqref{eq:subsys1} and~\eqref{eq:subsys3}.}
	\label{fig:StC}
\end{figure}

We collect $\mathcal N_i = {18816}$ samples, upon which we compute ${\theta_i^0}={\theta_i^a}=\theta_i = {0.0217}$. We fix {$\lambda_i = 0.1$}, and by solving SCP~\eqref{SCP}, we obtain the following optimal values and decision variables: 
\begin{align}\label{eq:platoon opt}
	&\mathds B_i(\vartheta_i^*, x_i) =0.1\,x_{1_i}+x_{2_i},\\\notag
	 &\mathcal S_i^{11^\ast}\! = 10^{-4},~ \mathcal S_i^{12^\ast} \!= [10^{-4}~~10^{-4}],~ \mathcal S_i^{21^\ast} \!= \mathcal S_i^{{12^\ast}^\top}\!\!, \\\notag
	&\mathcal S_i^{22^\ast} \!= \begin{bmatrix}
		10^{-4} & 10^{-4}\\
		10^{-4} & 10^{-4}
	\end{bmatrix}\!\!,~\sigma_i^\ast = 1.27, ~ \phi_i^\ast = 1.41,\\\notag
	 & {c_i^* = 1.07},~\beta_{\mathcal N_i}^\ast \!= 0.0018,~ \eta_{\mathcal N_i}^\ast = -0.1837.
\end{align}
Considering $\mu_i = 2^{-i}$, which satisfies Assumption~\ref{asmp:mu} with $\sum_{i\in\mathbb N^+}\mu_i=\sum_{i=1}^{\infty}2^{-i}=1$, as the subsystems are identical, we have $\sigma = \sum_{i\in \N^+}\mu_i\sigma_i^* = \sigma_i^*\sum_{i\in \N^+}\mu_i = 1.27$, and $\phi = \sum_{i\in \N^+}\mu_i \phi_i^*=\phi_i^*\sum_{i\in \N^+}\mu_i=1.41$, ensuring the fulfillment of condition~\eqref{Con0}. Moreover, we have $c= \sum_{i\in \N^+}\mu_ic_i^* = c_i^*\sum_{i\in \N^+}\mu_i = 1.07$ and $\lambda = \sup_{i\in\N^+}\{\lambda_i\} = 0.1$, satisfying condition~\eqref{Con0-1}. We now compute $\mathcal{L}_i^1 = 1.0050$ and $\mathcal{L}_i^2 = 0.1418$ according to Algorithm~\ref{alg:1}. As $\eta_{\mathcal N_i}^* + \mathcal{L}^1_{i} {\bar\theta_{i}}={-0.1619}\leq 0$ and $\eta_{\mathcal N_i}^* + \beta_{\mathcal N_i}^* + \mathcal{L}^2_{i} \theta_{i} = {-0.1788} \leq 0$, for all $i \in \mathbb{N}^+$, according to Theorem~\ref{Thm1}, $\mathds B(\vartheta,x)= \sum_{i \in \mathbb{N}^+}{\mu_i}\mathds B_i(\vartheta_i^*,x_i)$, with $\mathds B_i(\vartheta_i^*,x_i)$ as in \eqref{eq:platoon opt}, is a BC for the unknown infinite network with the unknown interconnection topology. For Algorithm~\ref{alg:1}, the runtime values are $T_{\text{Alg.1},1} \approx 0.0053$, $T_{\text{Alg.1},2} \approx 0.0536$, and $T_{\text{Alg.1},3} \approx 0.0032$, where $T_{\text{Alg.1},2}$ already includes $T_{\text{Alg.1},1}$. Thus, the total runtime is $T_{\text{Alg.1}} \approx T_{\text{Alg.1},2}+T_{\text{Alg.1},3}\approx 0.0568$ seconds for each subsystem. Moreover, the first phase of Algorithm~\ref{alg:2} takes $T_{\text{Alg.2}, 1} \approx {0.0132}$ seconds to execute, while the solver takes $T_{\text{Alg.2}, 2} \approx {0.0324}$ seconds to solve the SCP in~\eqref{SCP}, with a total runtime  $T_{\text{Alg.2}} \approx {0.0456}$ seconds per subsystem.
The satisfaction of conditions~\eqref{eq:subsys1}--\eqref{eq:csbceq} is shown in Figs.~\ref{fig:platoon-res} and \ref{fig:StC}. It is worth noting that while we assumed that all subsystems here are identical for the sake of simplicity, our compositional data-driven results are general and can handle heterogeneous infinite networks with finitely many subsystem classes.

\section{Conclusion}
We developed a data-driven approach within a compositional framework for constructing safety certificates for infinite networks, composed of a countably infinite number of subsystems, with both unknown models and interconnection topologies. The key message in our proposed infinite-network setting is that, while existing approaches exhibit an exponential growth in sample complexity with respect to the network size, our compositional strategy notably reduces this dependence by making it scale only with the effective dimension at the subsystem level. Furthermore, our compositional data-driven reasoning eliminates the need for the traditional dissipativity condition, which typically requires precise knowledge of the interconnection topology. 

\bibliographystyle{agsm}
\bibliography{biblio}

\begin{authorbio}[Mahdieh]{Mahdieh Zaker} received her B.Sc. from K. N. Toosi University of Technology, Tehran, Iran, in 2019, and her M.Sc. from Amirkabir University of Technology (Tehran Polytechnic), Tehran, Iran, both in Electrical Engineering, control major. She is currently a PhD student in the School of Computing at Newcastle University, UK. She is the Best Repeatability Prize Finalist at the $28^{\text{th}}$ ACM International Conference on Hybrid Systems: Computation and Control (HSCC), 2025. Her research interests are (nonlinear) control and systems theory, data-driven techniques, large-scale systems, and formal methods.\vspace{-0.7cm}
\end{authorbio}

\begin{authorbio}[Amy]{Amy Nejati} is an Assistant Professor in the School of Computing at Newcastle University in the United Kingdom. Prior to this, she was a Postdoctoral Associate at the Max Planck Institute for Software Systems in Germany from July 2023 to May 2024. She also served as a Senior Researcher in the Computer Science Department at the Ludwig Maximilian University of Munich (LMU) from November 2022 to June 2023. She received a PhD in Electrical Engineering from the Technical University of Munich (TUM) in 2023. She has received the B.Sc. and M.Sc. degrees both in Electrical Engineering. Her research mainly focuses on developing efficient (data-driven) techniques to design and control highly-reliable autonomous systems while providing mathematical guarantees. She was named a CPS Rising Star 2024, and her paper was selected as a Best Repeatability Prize Finalist at ACM HSCC 2025.\vspace{-0.5cm}
\end{authorbio}

\begin{authorbio}[Abolfazl]{Abolfazl Lavaei} is an Assistant Professor in the School of Computing at Newcastle University, United Kingdom. Between January 2021 and July 2022, he was a Postdoctoral Associate in the Institute for Dynamic Systems and Control at ETH Zurich, Switzerland. He was also a Postdoctoral Researcher in the Department of Computer Science at LMU Munich, Germany, between November 2019 and January 2021. He received the Ph.D. degree in Electrical Engineering from the Technical University of Munich (TUM), Germany, in 2019. He obtained the M.Sc. degree in Aerospace Engineering with specialization in Flight Dynamics and Control from the University of Tehran (UT), Iran, in 2014. He is the recipient of several international awards in the acknowledgment of his work including  Best Repeatability Prize (Finalist) at the ACM HSCC 2025, IFAC ADHS 2024, and IFAC ADHS 2021, HSCC Best Demo/Poster Awards 2022 and 2020, IFAC Young Author Award Finalist 2019, and Best Graduate Student Award 2014 at University of Tehran with the full GPA (20/20). His research interests revolve around the intersection of Control Theory, Formal Methods, and Statistical Learning Theory.
\end{authorbio}

\end{document}